\documentclass[usenatbib]{mn2e}

\usepackage{lscape,graphicx}

\newcommand{\microns}{$\mu$m}

\def\brg{Br\,$\gamma$}
\def\Pa{P\,$\alpha$}
\def\hei{He\,{\sc i}}
\def\heii{He\,{\sc ii}}
\def\civ{C\,{\sc iv}}
\def\niii{N\,{\sc iii}}
\def\ciii{C\,{\sc iii}}
\def\oiii{O\,{\sc iii}}

\def\ga{\mathrel{\hbox{\rlap{\hbox{\lower4pt\hbox{$\sim$}}}\hbox{$>$}}}}
\def\la{\mathrel{\hbox{\rlap{\hbox{\lower4pt\hbox{$\sim$}}}\hbox{$<$}}}}

\def\msun{M\mbox{$_{\normalsize\odot}$}}

\def\kms{\,km~s$^{-1}$}

\def\arcsec{$^{\prime \prime}$}
\def\arcmin{$^{\prime}$}

\def\hii{H{\sc ii}}

\newcommand{\fig}[1]{Fig.\ \ref{#1}}

\title[The young star-clusters Danks~1 and Danks~2]{The G305 star-forming
  complex: the central star clusters Danks~1 and
  Danks~2}
\author[B. Davies et al.]
{Ben Davies$^{1,2,3}$, J.~S. Clark$^{4}$, Christine Trombley$^{3}$, Donald
  F. Figer$^{3}$, \newauthor 
Francisco Najarro$^{5}$,  Paul A.\ Crowther$^{6}$, Rolf-Peter Kudritzki$^{7,8}$, Mark
  Thompson$^{9}$, \newauthor James S.\ Urquhart$^{10}$, Luke Hindson$^{9}$ \\
$^{1}$Institute of Astronomy, University of Cambridge, Madingley Road,
  Cambridge CB3 0HA, UK\\
$^{2}$School of Physics \& Astronomy, University of Leeds, Woodhouse
  Lane, Leeds LS2 9JT, UK\\
$^{3}$Center for Detectors, Rochester Institute of Technology, 54
  Memorial Drive, Rochester, NY 14623, USA\\
$^{4}$Department of Physics and Astronomy, The Open University, Walton
  Hall, Milton Keynes MK7 6AA, UK\\
$^{5}$Centro de Astrobiología (CSIC-INTA), Ctra. de Torrejón a Ajalvir km-4, 28850, Torrejón de Ardoz, Madrid, Spain\\
$^{6}$Department of Physics and Astronomy, University of Sheffield, Hounsfield Road, Sheffield, S3 7RH, UK\\
$^{7}$Institute for Astronomy, University of Hawaii, 2680 Woodlawn
  Drive, Honolulu, HI 96822, USA\\
$^{8}$Max-Planck Institute for Astrophysics,
  Karl-Schwarzchild-Str.\ 1,85748 Garching, Germany \\
$^{9}$Centre for Astrophysics Research, STRI, University of Hertfordshire, College Lane,
  Hatfield AL10 9AB, UK\\
$^{10}$Australia Telescope National Facility, CSIRO Astronomy and Space Science, PO Box 76, Epping, NSW 1710, Australia}
\begin{document}

\date{Accepted ... Received ...}

\pagerange{\pageref{firstpage}--\pageref{lastpage}} \pubyear{2009}

\maketitle

\label{firstpage}

\begin{abstract}
The G305 H\,{\sc ii} complex (G305.4+0.1) is one of the most massive
star forming structures yet identified within the Galaxy. It is host
to many massive stars at all stages of formation and evolution, from
embedded molecular cores to post main-sequence stars. Here, we present
a detailed near-infrared analysis of the two central star clusters
Danks~1 and Danks~2, using HST+NICMOS imaging and VLT+ISAAC
spectroscopy. We find that the spectro-photometric distance to the
clusters is consistent with the kinematic distance to the G305
complex, an average of all measurements giving a distance of
3.8$\pm$0.6kpc. From analysis of the stellar populations and the
pre-main-sequence stars we find that Danks~2 is the elder of the two
clusters, with an age of $3^{+3}_{-1}$Myr. Danks~1 is clearly younger
with an age of $1.5^{+1.5}_{-0.5}$Myr, and is dominated by three very
luminous H-rich Wolf-Rayet stars which may have masses $\ga$
100\msun. The two clusters have mass functions consistent with the
Salpeter slope, and total cluster masses of 8000$\pm$1500\msun\ and
3000$\pm$800\msun\ for Danks~1 and Danks~2 respectively. Danks~1 is
significantly the more compact cluster of the two, and is one of the
densest clusters in the Galaxy with $\log (\rho/M_{\odot}{\rm
  pc}^{-3}) = 5.5^{+0.5}_{-0.4}$. In addition to the clusters, there
is a population of apparently isolated Wolf-Rayet stars within the
molecular cloud's cavity. Our results suggest that the star-forming
history of G305 began with the formation of Danks~2, and subsequently
Danks~1, with the origin of the diffuse evolved population currently
uncertain. Together, the massive stars at the centre of the G305
region appear to be clearing away what is left of the natal cloud,
triggering a further generation of star formation at the cloud's
periphery.

\end{abstract}


\section{Introduction}
Massive stars have a profound effect on their wider Galactic
environment, via the production of copious quantities of ionising
radiation, and from the input of mechancial energy and chemically
processed matter into the interstellar medium (ISM). For these
reasons, an understanding of their lifecycle is of importance to many
areas of astronomy. Unfortunately, a number of questions regarding
this still remain unanswered, with the nature of their formation
mechansism(s) being particularly opaque. While growing observational
evidence suggests that stars between 20-40M$_{\odot}$ may form via disc
mediated accretion -- in a manner analagous to their lower mass
counterparts \citep[e.g. W33A \& W51N, see ][]{W33Apaper,Zapata09} -
it is still not clear how more massive stars form \citep{Z-Y07};
despite compelling observational evidence for stars with masses
significantly in excess of 40M$_{\odot}$ \citep[WR20a, NGC 3603-A1,
  and R145, ][]{Bonanos04,Rauw05,Schnurr08,Schnurr09}.

%Fortunately, given the difficulties in directly observing massive
%stellar formation in {\em action} - due to the likely brevity of the
%process, their intrinsic rarity and heavy extinction resulting from
%their natal molecular clouds - other indirect diagnotics of the
%process are available. One observational constraint, emboddied by the
%stars listed above, is the high binary fraction which appears to
%characterise such objects (e.g. Bosch et al. \cite{bosch}, Clark et
%al. \cite{clark08},\cite{clark09}, Ritchie et al. \cite{ritchie}). In
%principal this finding provides information on the nature of the
%accretion process, with differing scenarios resulting in different
%binary populations (e.g.  Clark \& Bonnell \cite{cb}).

%Extending this approach leads to the conclusion that sellar
Stellar hierarchies appear to be a signature of star formation, with
stars predominantly forming in clusters and, in turn, clusters forming
in larger complexes \citep[e.g.][]{Larsen04}. Such structures,
spanning tens to hundreds of parsecs, are most readily identifiable in
external star-forming galaxies such as M51 \citep[][]{Bastian05}. While
ages for individual clusters within the M51 complexes are difficult to
determine, it appears that they are likely to be rather youthful
(e.g. $<$10~Myr) and massive (3-30$\times$10$^4$M$_{\odot}$).
%with star formation rates (per unit area) consistent with {\em bona
%  fide} starburst galaxies. Indeed, such a connection is emphasised by
%the presence of similar complexes within interacting, starbursting
Similar complexes are seen within interacting, starbursting galaxies
such as the Antennae, where star formation rates are an order of a
magnitude higher than in M51, in turn yielding individual clusters
with masses $>$10$^6$M$_{\odot}$ \citep{Bastian05,Bastian06}.

%Given these properties,
A precise understanding of the nature of such complexes would be
invaluable for the following reasons: (i) they appear to represent a
ubiquitous mode of star formation in starburst galaxies; and (ii) by
virtue of their masses they provide a statistically well-sampled
stellar mass function.
% enabling us to identify both short lived
%pre-Main Sequence evolutionary phases and also obtain robust estimates
%of binary fractions. 
Unfortunately, the distances of their host galaxies and compact nature
conspire to make the determination of the properties of individual
clusters -- let alone stars -- observationally challenging. Therefore,
one might ask whether such structures are present within our own
Galaxy?  The recent detection of a number of massive
($\ga$10$^4$M$_{\odot}$) Red Supergiant (RSG) dominated clusters at the
base of the Scutum-Crux arm is {\em suggestive} of such a complex
\citep{Figer06,RSGC2paper,RSGC3paper,Negueruela10,Negueruela11}, although
their spatial extent ($\sim$100pc) and age spread ($\sim$10-20~Myr)
currently preclude an {\em unambiguous} association with a single,
physically distinct structure.

One observational approach to overcome such uncertainties is to search
for young, massive clusters still embedded in their natal giant
molecular cloud (GMC) and/or associated giant H\,{\sc ii} region.
Such a strategy guarantees the youth of such a complex, potentially
enabling individual examples of massive young stellar objects (MYSOs)
to be identified, and ultimately the global star formation history
from a spatially resolved census of the (proto-) stellar
populations. The latter goal is particularly important, since the
limited spatial resolution of such objects in external galaxies
precludes a detailed analysis of the processes by which the GMC is
converted into stars and star clusters. 

A number of GMCs which appear to contain both massive
($>$10$^3$~M$_{\odot}$), young clusters as well as deeply embedded
MYSOs have been identified \citep[e.g. W49A, W51 and the Carina
  nebula; ][]{Alves-Homeier03,Kumar04,Smith-Brooks07}. Another such
region is the G305 star forming complex \citep[$l$=305.4, $b$=+0.1;
]{C-P04}. Located in the Scutum-Crux arm at an estimated distance of
$\sim4$~kpc, it has the form of a large tri-lobed cavity with a
maximum extent of $\sim$34pc, delineated by both mid- and far-IR,
sub-millimetre and radio emission and centred on the young
clusters Danks 1 \& 2 (\fig{fig:widefield}). A large contingent of massive stars is
inferred from the ionising radiation required to support the total
radio flux (equivalent to the output from $>$30 canonical O7 V
stars). In addition, there are numerous signposts of ongoing star
formation in the cloud's periphery, in the form of deeply embedded
MYSOs, compact \hii-regions, and methanol and water masers
\citep{Urquhart07,Urquhart09,Hindson10,Clark11}. Finally, far-IR and
sub-millimetre continuum observations reveal the presence of a
significant reservoir of cold molecular material
($>$10$^5$M$_{\odot}$) available to fuel further star-forming activity
\citep[][ Clark et al. in prep]{Hindson10}.

% while the integrated lumnosity of a
%number of deeply embedded sources, coupled with the presence of a
%number of ucH\,{\sc ii} regions and methanol and water masers argues
%for a further generation of star formation on the periphery of the
%complex (Hindson et al. \cite{hindson}, Clark et
%al. \cite{clark11}). Indeed, IR spectra of members of a cluster
%located within a subordinate, compact H\,{\sc ii} region on the cavity
%perimeter reveals the presence of several additional mid-O - early-B
%stars (Leistra et al. \cite{leistra}); moreover far-IR and
%submm-continuum observations reveal the presence of a significant
%reservoir of cold molecular material ($>$10$^5$M$_{\odot}$) to fuel
%such activity (Hindson et al. \cite{hindson}, Clark et al. in prep.).

The overal morphology of the complex is strongly indicative of a
number of epochs of sequential star formation, initiated and sustained
by the action of the two central clusters. In this paper, we present a
near-IR analysis of these clusters, using high-resolution photometry
and spectroscopy, in order to determine their masses and ages, and
consequently whether their properties are consistent with such an
hypothesis.  We discuss and present these data in Section 2, report
our analysis in Section 3 and discuss these result both in the context
of the G305 complex and also in comparison to other star forming
regions in Section 4, before summarising our conclusions in Sect. 5.

\section{Observations \& data reduction}

\subsection{Photometry}

\subsubsection{Observations}
\begin{figure*}
  \centering
  \includegraphics[width=18cm,bb=5 20 795 566]{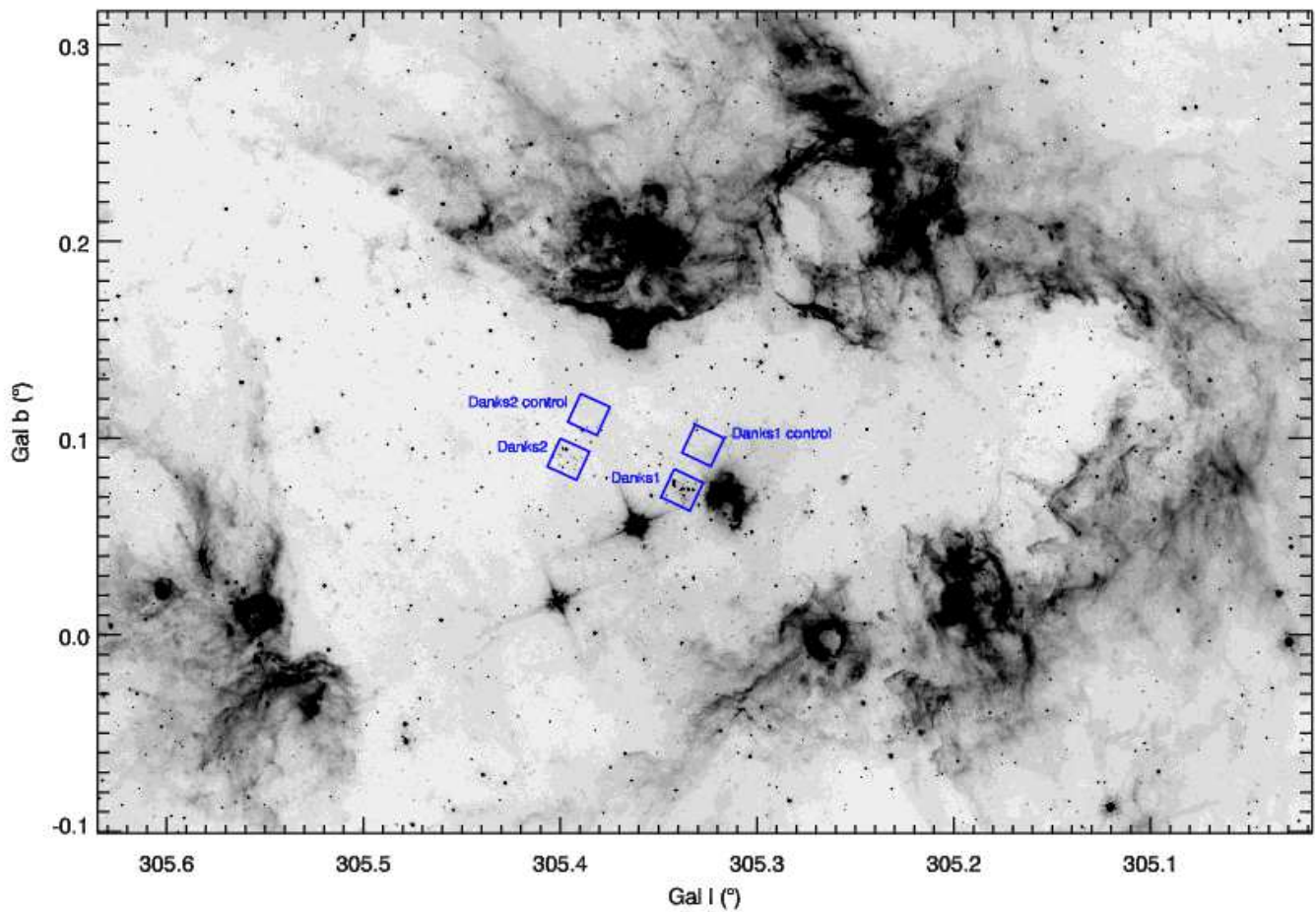}
  \caption{Wide-field Spitzer/GLIMPSE image of the G305 complex at
    5.8\microns, with the NICMOS fields overplotted. The
    5.8\micron\ band traces the PAH emission in the complex, and
    therefore the interface between ionised and molecular gas.  }
  \label{fig:widefield}
\end{figure*}

Images of the two clusters were obtained with HST/NICMOS on 16 July
2008. We used the NIC3 camera, which has a field-of-view of
51.2\arcsec$\times$51.2\arcsec\ and a pixel scale of 0.2\arcsec. The
clusters were imaged through each of the filters F160W and F222M, as
well as the narrow-band filters F187N and F190N which are centred on
P$\alpha$ and the neighbouring continuum respectively. In addition to
the clusters, in order to characterize the foreground population we
also imaged nearby control fields through the F160W and F222M
filters. The observed fields are indicated in
Fig.~\ref{fig:widefield}.

In our observations we employed a spiral dither pattern with six
separate pointings, each offset from the last by 5.07\arcsec. By
employing sub-pixel dithering we minimise the effects of non-uniform
intra-pixel sensitivity. We used the MULTIACCUM read mode, using
read-sequences and patterns that provided good sampling coverage over
a large dynamic range. The sampling sequences and total integration
times we used for each filter are listed in Table \ref{tab:samp}.

\begin{table}
  \centering
  \caption{Read-sequences and total integration times employed for each filter
    during the NICMOS observations. }
  \begin{tabular}{lccccc}
    \hline
    Filter & SAMP-SEQ & NSAMP & $T_{\rm int}$ (s) \\
    \hline
    \hline
    F160W  & STEP2    & 17    &  168 \\
    F222M  & STEP8    & 12    &  448 \\
    F187N  & STEP8    & 10    &  240 \\
    F190N  & STEP8    & 10    & 240  \\
    \hline
  \end{tabular}
  \label{tab:samp}
\end{table}

\subsubsection{Data reduction and analysis}
\begin{figure*}
  \centering
  \includegraphics[width=8.5cm,bb=0 0 640 623]{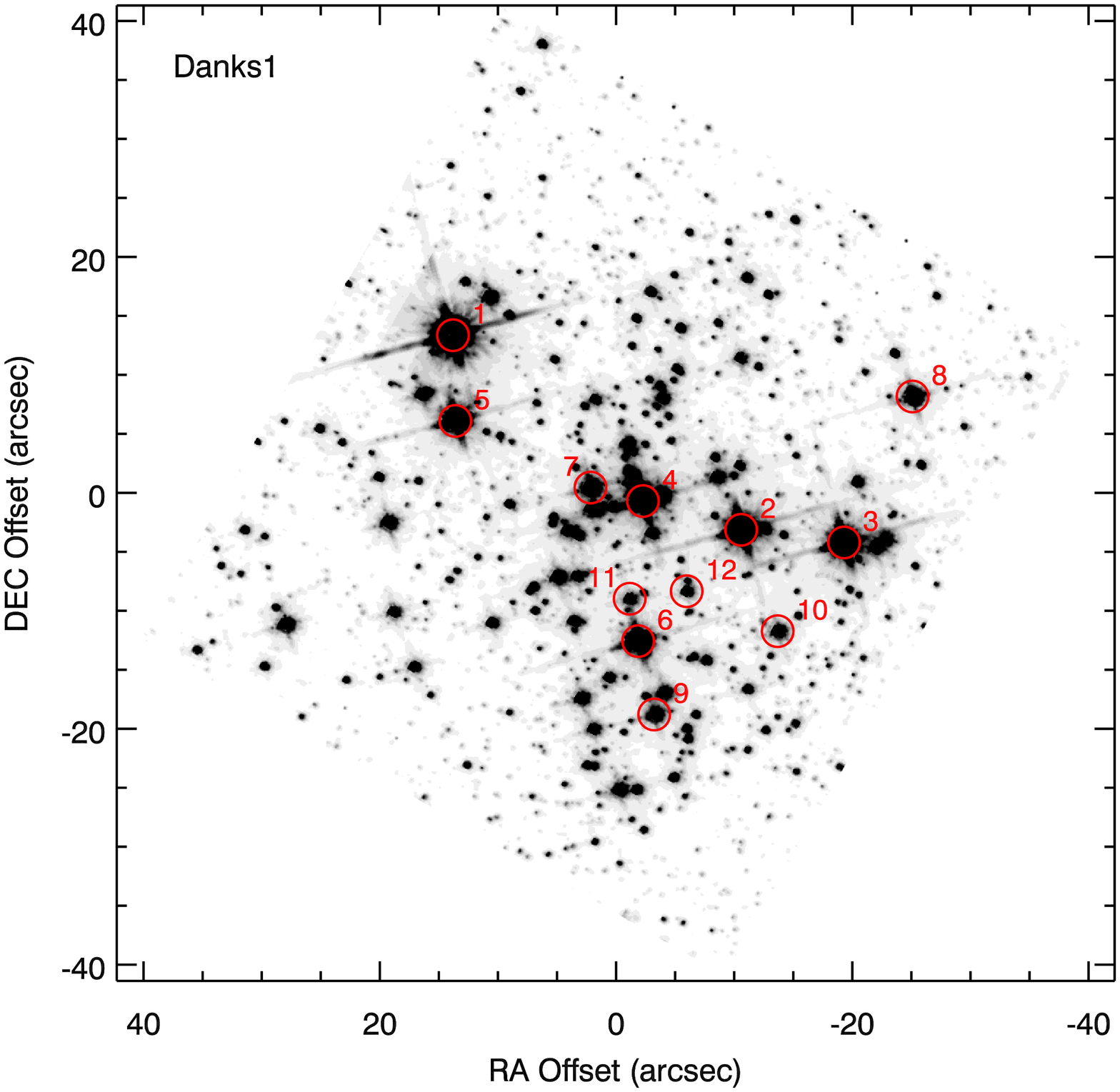}
  \includegraphics[width=8.5cm,bb=0 0 640 623]{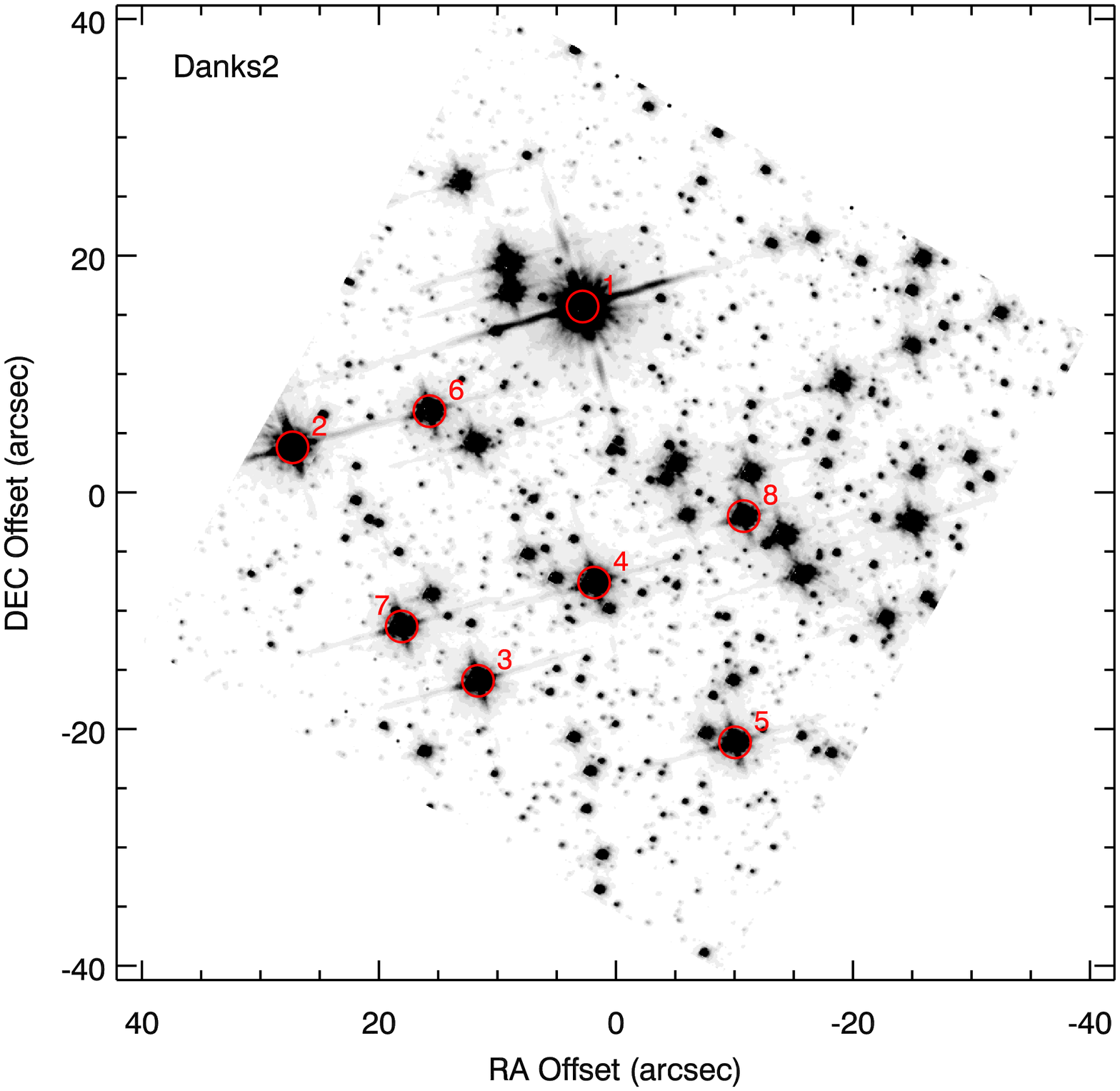}
  \caption{F160W mosaics of each cluster; {\it left}: Danks~1, {\it
      right}: Danks~2. In each image north is up and east is left. The
    stellar identifications (see Table \ref{tab:spectypes}) are denoted
    by the red circles. }
  \label{fig:mosaic}
\end{figure*}

Following the guidelines of the NICMOS Data Handbook v7.0 we performed
the standard reduction steps of bias subtraction, dark-current
correction and flat-fielding using the custom-written NICMOS software
{\sc calnica}. as our dithered observations made three sub-samples of each
pixel we re-sampled each dithered observation onto a finer grid
containing a factor of 3 more pseudo-pixels. The six images were then
mosaicked together, again using the custom-written NICMOS
software. The final mosaics of each cluster are shown in
Fig.\ \ref{fig:mosaic}. 

In extracting the photometry from the images, we experimented with
several methods and algorithms. We found that the best results
were obtained (fewest spurious source detections, most effective
deblending of sources) using the {\sc starfinder} routines
\citep{STARFINDER} which run under the IDL environment. The algorithm
uses the image point-spread function (PSF) in order to locate
point-sources. We found that the algorithm was most effective when
using the synthetic PSF generated by {\sc tinytim}. To check for
consistency between this method and regular aperture photometry, we
ran both {\sc starfinder} and {\sc daophot} on the relatively
uncrowded control fields and found no systematic differences between
the two.

In order to characterize the statistical completeness in each field
observed, as well as make empirical measurements of the photometric
errors, we performed artificial star experiments on both cluster
fields. We used {\sc tinytim} to generate the artificial stars with a
luminosity function identical to that observed in each field. We then
randomly sampled artificial stars from this luminosity function and
added them to the image. No more than 100 stars were added to the
image so as not to significantly alter the level of crowding. The
photometry algorithm was then re-run on the artificial image, and the
output photometry and astrometry compared with the locations and
magnitudes of the input stars. For a star to be considered `recovered'
we specified a maximum distance separation between input position and
detected position of 0.22\arcsec, or roughly 1$\sigma$ of the
PSF. Additionally, if the output magnitude of a star was more than a
factor of 2 greater than the input magnitude then it was assumed that
that star had been blended with a brighter one, and that the input
star was lost. The 50\% completeness limits were found to be at
$m_{160W} = 18.8$ and $m_{222M} = 17.4$ for Danks~1, and $m_{160W} =
19.2$ and $m_{222M} = 18.3$ for Danks~2. The slightly fainter
detection limit for Danks~2 with respect to Danks~1 is due to the
reduced level of crowding in that field. 

% Figures \ref{fig:comp1} and \ref{fig:comp2} illustrate
%the photometric precision and completeness as a function of input
%magnitude for both clusters through each filter.

\subsection{Spectroscopy}

\subsubsection{Observations}
Spectra were obtained of several stars in each cluster using the ESO
ISAAC instrument mounted on UT1 of the VLT\footnote{ESO programme ID
  077.C-0207(B), PI J.S.~Clark}. Observations were taken in good
weather on the nights of June 27th to July 2nd 2006. The instrument
was used in medium resolution mode with the 0.3\arcsec\ slit, and two
overlapping wavelength settings were used per target to achieve a
spectral range of 2.04-2.21\microns\ at a spectral resolution of
$\lambda/\Delta\lambda$=8900.

The observing strategy was to align the slit in order to obtain
spectra of two programme stars simultaneously. The stars were nodded
along the slit in an ABBA pattern. Integration times per slit position
were limited to $\sim$100 seconds to allow for accurate subtraction of
the sky emission, with each integration split into 4-10 seperate
read-outs to avoid saturation around strong stellar emission
lines. The total integration times per star were between 400-500
seconds. The star HD~113457 (spectral type A0\,{\sc v}) was
observed after every two target observations as a measure of the
atmospheric absorption. Continuum lamp exposures were taken just as
frequently in order to correct for any fringing on the detector.

\subsubsection{Data reduction}
Nod pairs were subtracted from one another to remove bias level, dark
current and sky emission lines. Before extracting the spectra, the
frames were corrected for the degree of warping which is present in
ISAAC spectroscopic data. Warping in the spatial direction was
characterized by fitting polynomials to the stellar spectral traces
across the detector. To measure the warping in the dispersion
direction a `sky lines' image was created by summing all nod pairs of
the same target and subtracting the stars. The sky lines were then
fitted with polynomials to wavelength-calibrate the data and to
measure the degree of warping in the dispersion direction. Using the
fits to the stellar traces and the sky lines each frame was then
resampled onto a linear, wavelength calibrated grid. The wavelength
solution of each frame had an r.m.s.\ of 0.2-0.3 pixels, or 3-6\kms. 

In some observations, it was noticed that a degree of diffuse
line-emission was present due to ionized nebulae in the vicinity of
the clusters. This emission was subtracted from each frame by fitting
gaussian profiles to the spatial variations and interpolating across
any stellar traces that it intersected. 

\begin{figure*}
  \centering
  \includegraphics[height=22.5cm]{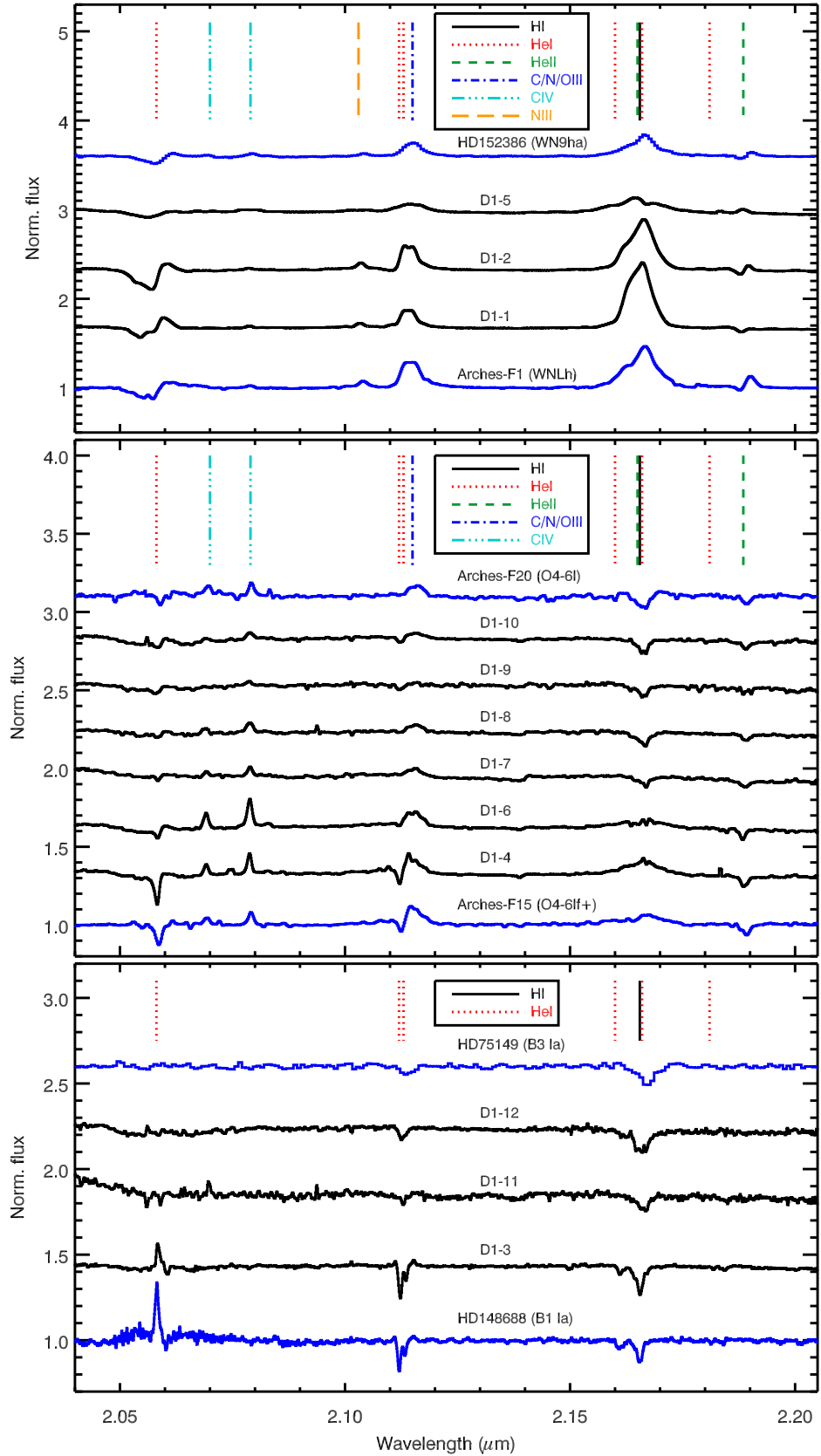}
  \caption{Spectra of the stars in Danks~1. Also shown are comparison
    spectra of template stars taken from \citet{Hanson96,Hanson05},
    \citet{Martins08}, \citet{Figer97}, and \citet{W-H97}. The
    locations of key diagnostic spectral lines have been
    indicated. The three panels show the WNh stars ({\it top}), the
    early-mid O stars ({\it middle}), and the late-O / early-B stars
    ({\it bottom}).  }
  \label{fig:spec1}
\end{figure*}

\begin{figure*}
  \centering
  \includegraphics[height=22.5cm]{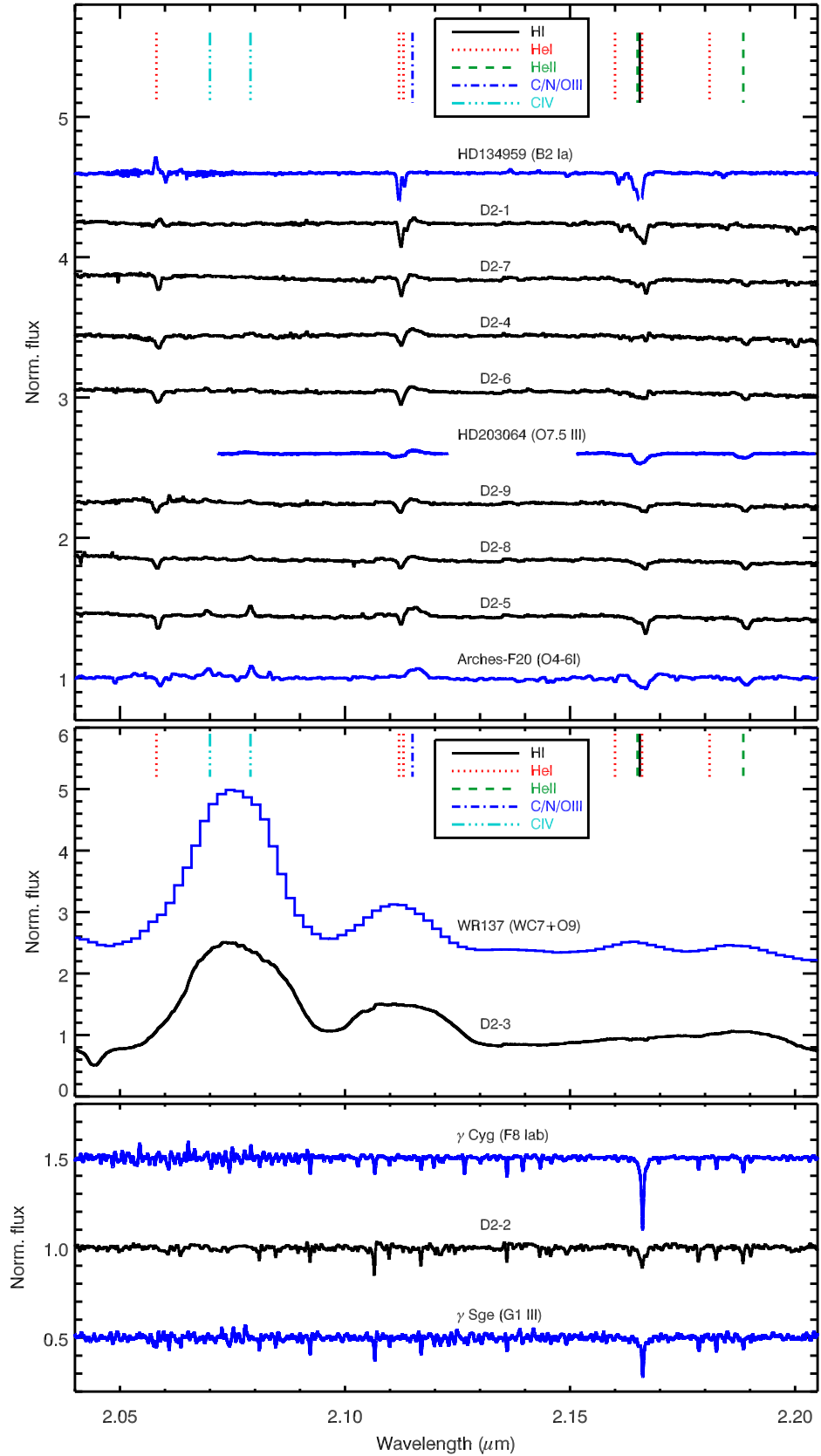}
  \caption{Same as \fig{fig:spec1} but for Danks~2. The three panels
    show the OB stars ({\it top}), the WCL star ({\it middle}), and
    the yellow star ({\it bottom}).}
  \label{fig:spec2}
\end{figure*}

Spectra were extracted by summing across the pixels around each
stellar trace. Cosmic ray hits and bad pixels were rejected by
comparing repeat observations of the same star. Before dividing
through by the telluric standard spectrum, we first fitted the
standard's \brg\ absorption feature using a Voigt profile, and
corrected for the continuum slope by dividing through by a black-body
appropriate for the star's spectral type. The standard star spectrum
was then cross-correlated with the science target to correct for any
small sub-pixel shifts which may produce artifacts in the final
spectrum. The signal-to-noise ratio (SNR) of the fully reduced target
spectra was estimated from flat regions of continuum, and was found to
be typically 150-400. The final spectra for the stars in Danks~1 and 2
are shown in Fig.\ \ref{fig:spec1} and \fig{fig:spec2} respectively.

%\begin{figure*}
%  \centering
%  \includegraphics[width=8.8cm]{D1-Ofpe.eps}
%  \includegraphics[width=8.8cm]{D2-Ofpe.eps}
%  \caption{Spectra of the stars in the two clusters. \textit{Left}:
%    Danks~1; \textit{Right}: Danks~2. Also shown are comparison
%    spectra of template stars taken from \citet{Hanson96,Hanson05},
%    \citet{Martins08}, \citet{Figer97}, and \citet{W-H97}. The
%    locations of key diagnostic spectral lines have been indicated.}
%  \label{fig:spec}
%\end{figure*}

\subsubsection{Spectral classification} \label{sec:spectypes}
With the exception of one object (D2-2), we found that all stars
observed exhibited spectral features associated with hot stars. The
diagnostic features observed were \brg, \hei\ (2.056\microns,
2.112\microns, 2.165\microns), \heii\ (2.189\microns), the blend of
\ciii, \niii\ and \oiii\ at 2.115\microns, \niii\ (2.103\microns), and
\civ\ (2.070\microns, 2.079\microns).

In order to classify the spectra of the stars, we used the works of
\citet{Hanson96,Hanson05}, \citet{Morris96}, \citet{Figer97} and
\citet{Bohannan-Crowther99}. Also, since a number of Danks stars have
$K$-band spectra which are remarkably similar to those in the Arches
cluster, we make comparisons to the spectra of those stars
presented in \citet{Martins08}. Given the comparative paucity of
emission features suitable for classification, spectral types
determined solely from K~band spectra are inevitably less precise than
those obtained from optical data. Nevertheless, our spectral
classification scheme is as follows:

\begin{itemize}
\item {\it WNLh}: broad emission lines of \brg, the
  2.115\microns\ complex, \civ\ and \niii. The \hei\ (2.056\microns)
  and \heii (2.189\microns) features have complex structures and/or
  P~Cygni-type profiles. Emission from \civ\ is either absent or very
  weak.

\item {\it O hypergiant (OIf)}: broad emission of \brg\ but with low
  contrast to the continuum; 2.115\microns\ complex and \civ\ in
  emission; \hei\ (2.056\microns, 2.112\microns) and
  \heii\ (2.189\microns) in absorption. 

\item {\it O-type}: broad \brg\ in absorption, though for
  supergiants it may be in emission. \civ\ emission and
  \heii\ (2.189\microns) absorption is seen in spectral types no later
  than O7 and O8 respectively. \hei\ (2.112\microns) is in absorption
  with the \hei\ 2.113\microns\ multiplet in emission for types later
  than O8. \hei\ (2.056\microns) is heavily dependent on the wind
  properties and may be in either emission or absorption. 

\item {\it late O-/early B-type}: \brg, \hei\ (2.112\microns) and
  \hei\ (2.162\microns) in absorption. Absence of \civ\ in
  spectral-types later than O8, and no \heii\ types later than O9. The
  \hei\ (2.112\microns) feature is absent in spectral-types later than
  $\sim$B5, while \hei\ (2.056\microns) is not seen beyond $\sim$B3 --
  though we note again that this line is very sensitive to wind
  density.
\end{itemize}

The other spectra which do not fall into these categories are easily
classifiable: D2-3 has the broad \civ\ and 2.115\microns\ emission
lines of a WC7/WC8 star (see Fig.\ \ref{fig:spec2}, centre panel), and
has been previously identified by \citet[][ classified by those
  authors as WC8]{Mauerhan09}; while D2-2 has the spectral features of
a late F-type star but without the dense molecular absorption of an
early G-type star (Fig.\ \ref{fig:spec2}, bottom panel). The near-IR
colour of this star suggests that it may be a foreground star.

The spectral types attributed to each star, along with their
coordinates and photometry, are listed in Table
\ref{tab:spectypes}. In Danks~1, we found three stars with spectral
features attributable to H-rich WR stars (type WNLh), stars D1-1, D1-2
and D1-5\footnote{\citet{Mauerhan11} also detected D1-1 and D1-5. They
  classified D1-5 as WN9h, consistent with our classification of
  WNLh. However, D1-5 was classified as WN9, whereas our high
  signal-to-noise spectrum detected weak emission from \brg\ and so
  was classified WNLh.}. Stars D1-4 and D1-6 have strong \civ\ in
emission as well as very broad \brg\ emission, while
\hei\ (2.112\microns) is in absorption. These stars are assigned
spectral types O6-8If\footnote{We note that \citet{Martins08} assigned
  types of O4-6If+ for stars with similar spectral appearance in the
  $K$-band. This is representitive of the uncertainty when classifying
  hot stars in this spectral window.}. Stars D1-7, D1-8, D1-9 have
weak \civ\ and \heii, with no \hei\ (2.112\microns), and are
classified as O4-6. D1-10 is similar to these stars except that
\hei\ (2.112\microns) is in absorption, implying a slightly later
spectral type of O6-8. Stars D1-11 and D1-12 have no \heii\ and weak
\hei\ 2.112\microns, and are classified as O8-B3. The width of the
absorption lines in the spectra of D1-10, D1-11 and D1-12 mean that
they are likely to have luminosity classes of {\sc v-iii}. Star D1-3
has very narrow absorption lines of \brg\ and
\hei\ (2.112,2.162\microns) while \hei\ 2.056\microns\ is in emission,
and so is considered to be a O8-B3 supergiant.

In Danks~2, aside from the WC and yellow star, we find 3 stars with
very weak \civ\ emission, weak \brg\ absorption and
\hei\ (2.112\microns) absorption (D2-5, D2-8, D2-9), which we classify
as O6-8. A further three stars (D2-4, D2-6, D2-7) are
spectroscopically similar, but have no \civ, and so have slightly
later types of O8-9. The remaining stars (D2-1 and D2-7) are
classified as O8-B3, with the former star having the narrower lines
and \hei\ 2.056\microns\ emission typical of a supergiant.

Finally, we remark that from analysis of the \Pa\ fluxes of the stars
in each field (see Sect.\ \ref{sec:anal}), it is unlikely that there
are any further strong emission-line stars in either cluster, aside
from those presented here. 

\begin{table*}
  \centering
  \caption{Coordinates (J2000), spectral types and photometry of the stars
    with spectroscopic data. $^{\dagger}$This photometry is taken from the 2MASS catalogue
  $H$-band data, as our F160W data was corrupted for this star. }
  \begin{tabular}{llllllllccccc}
    \hline \hline
    Star & Alt. ID & \multicolumn{3}{c}{RA} & \multicolumn{3}{c}{DEC} & 
    Spec Type & $m_{\rm F160}$ & $m_{\rm F222}$ & $m_{\rm F187}$ & $m_{\rm F190}$ \\    
\hline
{\it Danks 1} \\
 D1-1 & MDM 8 & 13$^h$ & 12$^m$ & 28.49$^s$ & -62\degr & 41\arcmin & 43.46\arcsec & WNLh   &  7.260$^{\dagger}$ &  6.620 &  6.094 &  7.002\\
 D1-2 & - & 13 & 12 & 24.95 & -62 & 41 & 59.92 & WNLh   &  8.158 &  7.460 &  6.941 &  7.752\\
 D1-3 & - & 13 & 12 & 23.69 & -62 & 42 &  0.99 & O8-B3I &  8.223 &  7.618 &  7.840 &  7.925\\
 D1-4 & - & 13 & 12 & 26.16 & -62 & 41 & 57.50 & O6-8If &  8.934 &  8.262 &  8.166 &  8.511\\
 D1-5 & MDM 7 & 13 & 12 & 28.47 & -62 & 41 & 50.72 & WNLh   &  8.834 &  8.306 &  7.943 &  8.518\\
 D1-6 & - & 13 & 12 & 26.22 & -62 & 42 &  9.37 & O6-8If &  9.220 &  8.647 &  8.595 &  8.857\\
 D1-7 & - & 13 & 12 & 26.80 & -62 & 41 & 56.36 & O4-6   & 10.161 &  9.540 &  9.685 &  9.754\\
 D1-8 & - & 13 & 12 & 22.84 & -62 & 41 & 48.60 & O4-6   & 10.355 &  9.685 &  9.817 &  9.905\\
 D1-9 & - & 13 & 12 & 26.02 & -62 & 42 & 15.59 & O4-6   & 10.806 & 10.239 & 10.368 & 10.398\\
D1-10 & - & 13 & 12 & 24.50 & -62 & 42 &  8.52 & O6-8   & 11.319 & 10.612 & 10.805 & 10.864\\
D1-11 & - & 13 & 12 & 26.32 & -62 & 42 &  5.78 & O8-B3  & 11.804 & 11.255 & 11.431 & 11.455\\
D1-12 & - & 13 & 12 & 25.62 & -62 & 42 &  5.13 & O8-B3  & 12.227 & 11.598 & 11.800 & 11.836\smallskip \\
{\it Danks 2} \\
 D2-1 & - & 13 & 12 & 56.34 & -62 & 40 & 27.78 & O8-B3I &  7.750 &  6.810 &  7.027 &  7.124\\
 D2-2 & - & 13 & 12 & 59.90 & -62 & 40 & 39.71 & F8-G1  &  9.131 &  8.912 &  8.935 &  8.997\\
 D2-3 & - & 13 & 12 & 57.63 & -62 & 40 & 59.42 & WC7-8  &  9.911 &  9.146 &  8.392 &  9.443\\
 D2-4 & - & 13 & 12 & 56.20 & -62 & 40 & 51.11 & O8-9   & 10.108 &  9.537 &  9.616 &  9.783\\
 D2-5 & - & 13 & 12 & 54.48 & -62 & 41 &  4.60 & O6-8   & 10.001 &  9.605 &  9.625 &  9.717\\
 D2-6 & - & 13 & 12 & 58.22 & -62 & 40 & 36.64 & O8-9   & 10.125 &  9.631 &  9.740 &  9.813\\
 D2-7 & - & 13 & 12 & 58.56 & -62 & 40 & 54.84 & O8-9   & 10.185 &  9.669 &  9.755 &  9.871\\
 D2-8 & - & 13 & 12 & 54.37 & -62 & 40 & 45.48 & O6-8   & 10.333 &  9.830 &  9.967 & 10.034\\
 D2-9 & - & 13 & 12 & 57.38 & -62 & 40 &  1.43 & O6-8   &      - &      - &      - &      -\smallskip \\
 \hline
  \end{tabular}
  \label{tab:spectypes}
\end{table*}

\begin{table*}
  \centering
  \caption{Coordinates (J2000), spectral types and 2MASS photometry of the
    other confirmed massive stars in the G305 complex, taken from the
    literature. References are: 1 -- \citet{Mauerhan11}; 2 --
    \citet{Shara09}; 3 -- \citet{vanderHucht01}; 4 --
    \citet{Leistra05}; 5 -- \citet{C-P04}.}
  \begin{tabular}{lcllllllcrrrc}
    \hline \hline
    ID & Alt. ID & \multicolumn{3}{c}{RA} & \multicolumn{3}{c}{DEC} & 
    Spec Type & $J$ & $H$ & $K_{S}$ & Ref. \\    
\hline
MDM 3       & -    & 13$^h$ & 12$^m$ & 09.05$^s$ & -62\degr & 43\arcmin & 26.7\arcsec & WN8-9 & 10.21 & 8.57  & 7.58  & 1 \\
S09 845-34 & MDM 4 & 13 & 12 & 21.30 & -62 & 40 & 12.5 & WC8   & 10.75 & 9.57  & 8.77  & 1,2 \\
MDM 5      & -     & 13 & 12 & 25.46 & -62 & 44 & 41.7 & WN9   & 9.81  & 8.48  & 7.65  & 1 \\
S09 845-35 & MDM 6 & 13 & 12 & 27.66 & -62 & 44 & 22.0 & WC7   & 13.16 & 11.82 & 10.71 & 1,2 \\
WR48a      & -     & 13 & 12 & 39.65 & -62 & 42 & 55.8 & WC6   & 8.74  & 6.80  & 5.09  & 3 \\
S09 847-8  & -     & 13 & 12 & 45.35 & -63 & 05 & 52.0 & WN6   & 13.06 & 11.34 & 10.26 & 2 \\
L05-A1     & -     & 13 & 11 & 41.04 & -62 & 32 & 50.8 & O5-6 \sc{i} & 11.75 & 10.39 & 9.58  & 4 \\
L05-A2     & -     & 13 & 11 & 33.88 & -62 & 33 & 27.1 & B0-1 \sc{v} & 12.31 & 11.02 & 10.34 & 4 \\
L05-A3     & -     & 13 & 11 & 39.50 & -62 & 33 & 28.2 & B2-3 \sc{v} & 14.06 & 12.65 & 11.97 & 4 \\
MSX305.4013+00.0170 & - & 13 & 13 & 02.04 & -62 & 45 & 03.3 & WCL & 6.59 & 5.00 & 3.95 & 5 \\
\hline
  \end{tabular}
  \label{tab:otherstars}
\end{table*}

\section{Results \& analysis} \label{sec:anal}
Our data have revealed a large number of massive stars within the two
central clusters. These are in addition to the numerous and apparently
isolated massive stars within G305 \citep{Shara09,Mauerhan11}. In
Table \ref{tab:spectypes} we list the astrometry, photometry and
spectral types of these stars. For completeness, in Table
\ref{tab:otherstars} we list the other known massive stars in the
vicinity of the two clusters.

\begin{figure*}
  \centering
  \includegraphics[width=17cm,bb=0 30 850 562]{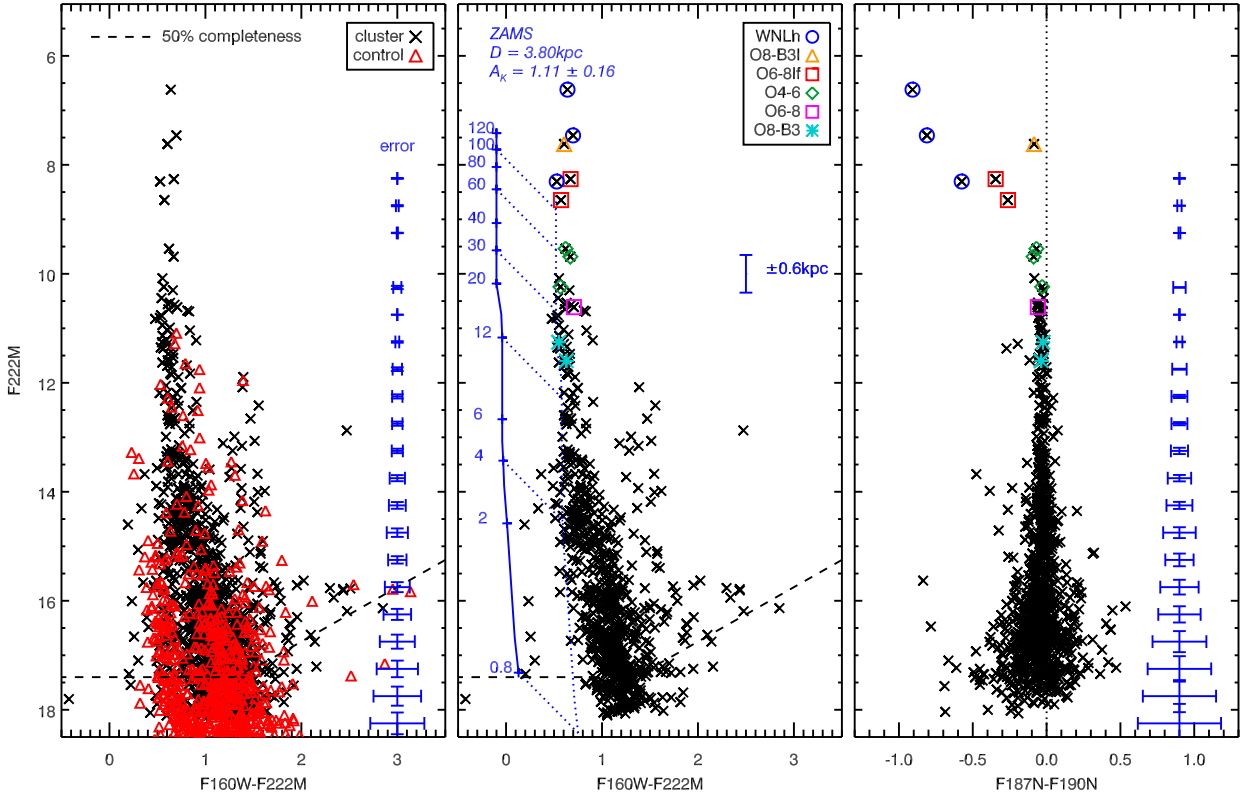}
  \caption{Photometry of Danks~1. The left panel shows the photometry
    of both the cluster and control fields, while the photometric
    errors and 50\% completeness level (dashed line) are also
    shown. The centre panel shows the photometry of the cluster after
    being decontaminated of field stars, and also indicates the stars
    with known spectral types. The right panel shows the \Pa-excess of
    the stars in the cluster field, with the stars with known spectral
    types again indicated. \vspace{-6mm}}
  \label{fig:phot1}
\end{figure*}

\begin{figure*}
  \centering
  \includegraphics[width=17cm,bb=0 30 850 562]{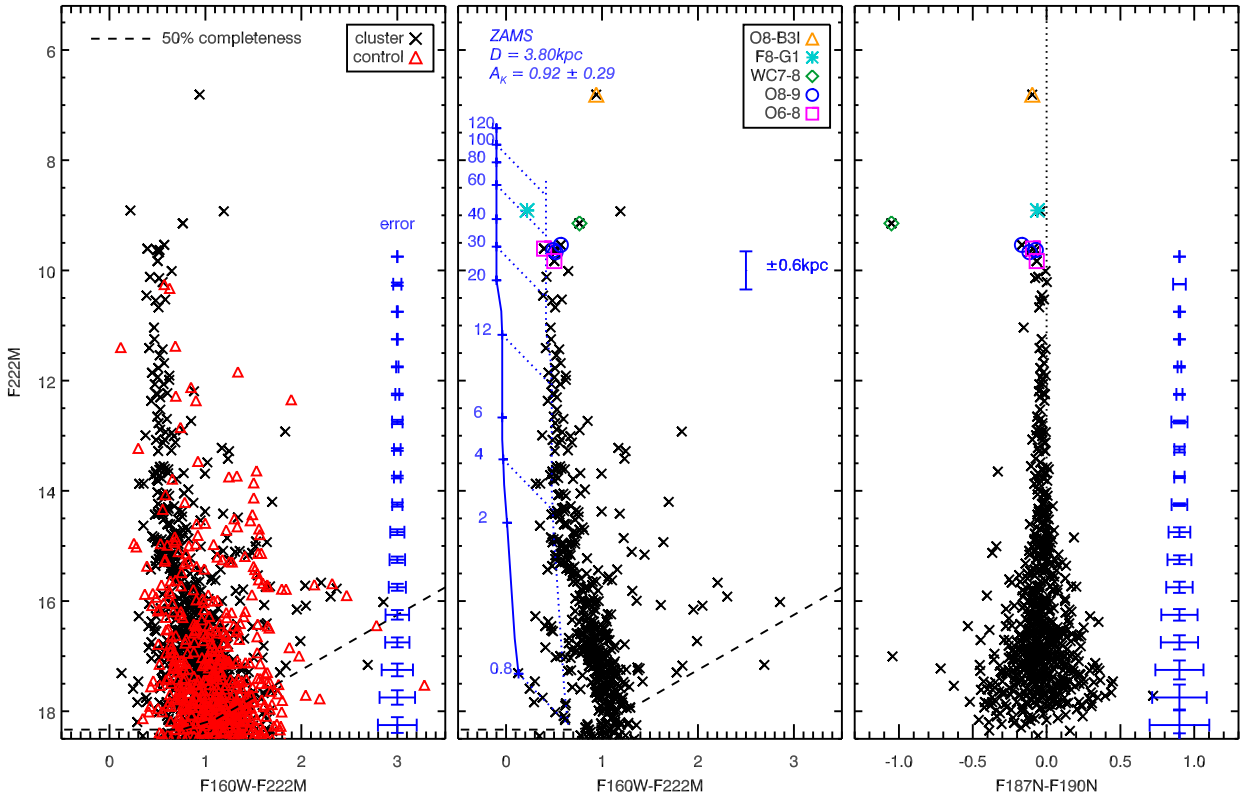}
  \caption{Same as Fig.\ \ref{fig:phot1} but for the field centred on
    Danks~2.}
  \label{fig:phot2}
\end{figure*}

In Figs \ref{fig:phot1} and \ref{fig:phot2} we plot the results of the
photometry for each cluster. In the left panel of each figure we show
the colour-magnitude diagrams (CMDs) of the cluster and control
fields. In the centre panel we have decontaminated the cluster field
of foreground stars using the control field: for every star in the
control field, we subtract a corresponding star in the cluster field
with $f_{\rm 222M}$ and $(f_{\rm 160W}-f_{\rm 222M})$ within the
photometric errors at that brightness. As the formal photometric
errors can be very small for bright objects, we specify a minimum
`errorbox' of size $f_{\rm 222M} = \pm 0.1$mags, $(f_{\rm 160W}-f_{\rm
  222M}) = \pm 0.14$mags for the decontamination algorithm. In the
centre panels we also illustrate the brightnesses and colours of
zero-age main sequence (ZAMS) stars at the distances of the clusters
(see Sect.\ \ref{sec:dist} for discussion on cluster
distances)\footnote{These ZAMS tracks were computed by first taking
  the masses, temperatures and luminosities of ZAMS stars from
  \citet{Mey-Mae00}. The relation between spectral type and
  temperature was taken from \citet{Martins05} for stars with masses
  $>$15\msun\ and \citet{Johnson66} for the rest. Infrared magnitudes
  and colours were taken from \citet{Martins-Plez06} and
  \citet{Koornneef83}.}. In the right-hand panel we show the
photometry across \Pa.

In the following sections we use the information in these plots, as
well as the spectral types of the stars, to determine the physical
properties of each cluster. In our analysis we will equate the NICMOS
filters F160W and F222M with the photometric bands $H$ and $K$
respectively.


\subsection{Extinction} \label{sec:extinct}
The extinction towards each cluster is derived from the brightest
stars in the decontaminated CMDs. The median F160W-F222M colour is
obtained from all stars in each cluster with F222M magnitudes brighter
than 13 whose locus is close to the ZAMS, since these stars show no
evidence for circumstellar extinction. As the intrinsic F160W-F222M
colours of hot stars are all approximately zero, we take the median
colour as a colour {\it excess}, and convert it into an extinction
using the law of \citet{R-L85}. We find average line-of-sight
extinctions for the two clusters of $A_{K} = 1.11\pm0.16$ and $A_{K} =
0.92\pm0.29$ for Danks~1 and Danks~2 respectively \footnote{These
  extinctions were computed using a value of $\alpha = -1.53$ for the
  slope of the IR extinction law. If a slope of $\alpha = -2.14$ is
  used \citep{Stead-Hoare09}, we find $A_{K} = 0.6\pm0.1$ and $A_{K} =
  0.6\pm0.2$ for Danks~1 and Danks~2 respectively. }. We adopt these
values throughout the rest of this work.

\subsection{Distances} \label{sec:dist}
In order to calculate the distance to the clusters, we employ two
independent and complementary methods. We use various radial
velocities measurements for the G305 complex to determine a kinematic
distance, and the photometry of stars in each cluster to calculate
spectro-photometric distances. 

\subsubsection{Kinematic distance to G305}
\begin{figure}
  \centering
  \includegraphics[width=8.5cm]{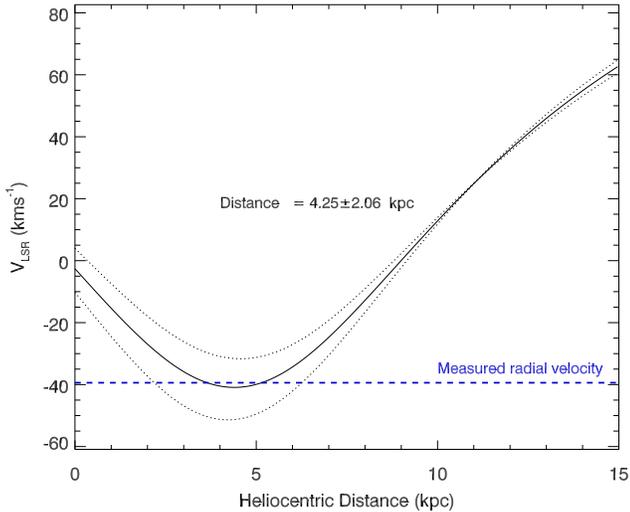}
  \caption{Galactic rotation curve in the direction of G305 (solid
    line). The dotted lines show the extremes obtained for the
    rotation curve when the uncertainties in Galactocentric distance
    and Solar rotation velocity are taken into account. }
  \label{fig:grotcurve}
\end{figure}

Many \hii-regions and young stellar objects (YSOs) are found in the
dense molecular material surrounding the two clusters, and for many of
these objects radial velocity measurements exist. A search of the {\it
  RMS} database \citep{Hoare05}\footnote{{\tt
    http://www.ast.leeds.ac.uk/RMS}} of YSOs in the region displayed
in Fig.\ \ref{fig:widefield} yielded 15 objects with known radial
velocities \citep[][ and refs therein]{Urquhart07,Urquhart09}. The
mean radial velocity and standard deviation is $v_{\rm LSR} = -39.4
\pm 3.0$\kms. In Fig.\ \ref{fig:grotcurve} we compare this value to
the Galactic rotation curve in the direction of G305. We use the
rotation curve of \citet{B-B93}, and the values for the
Galacto-centric distance (7.6$\pm$0.3\, kpc) and Solar angular
velocity (214$\pm$7\kms) compiled by \citet{K-D07}. It can be seen
from the plot that the average G305 radial velocity is close to the
tangential point. Once the uncertainties in the rotation curve are
taken into account, we find a distance of 4.2$\pm$2.0\,kpc.

\subsubsection{Spectrophotometric distances of the two clusters}
In calculating the spectrophotometric distances to each cluster, we
begin by analysing those stars for which we are confident of the
luminosity classes. As discussed in Sect.\ \ref{sec:spectypes}, the
star D1-3 shows the clear spectral signatures of a O8-B3
supergiant. From \citet{Crowther06}, such stars have absolute $K$-band
magnitudes of $M_{K}=-6.2 \pm 0.5$, and so for this star we find a
spectrophotometric distance to Danks~1 of
$3.48_{-0.71}^{+0.91}$kpc. Applying the same analysis to the star
D2-1, another O8-B3 supergiant, we find a distance to Danks~2 of
$2.61^{+0.82}_{-0.61}$kpc. Both these distances are within the errors
of one another, as well as being consistent with the kinematic
distance derived in the previous section. 

In Table \ref{tab:specphot} we list the spectrophotometric distances
for the remaining stars in each cluster which do not display evidence
for high luminosity classes (i.e.\ no \brg\ emission, indicative of
supergiants). Distances are calculated for each of the {\sc v}, {\sc
  iii} and {\sc i} luminosity classes, based on the absolute $K$-band
magnitudes quoted in the `observational' temperature scale of
\citet{Martins-Plez06}. By making the a priori assumption that the
stars with similar spectral types have the same luminosity class we
can assign luminosity classes that form a consistent picture of the
distances to each cluster.

In Danks~1, consistent distances are found for all the
`non-supergiants' if we assign them to be class {\sc v} objects,
i.e.\ on or near the main sequence. A weighted mean of these distances
gives $4.16\pm0.6$kpc, where the uncertainty is the r.m.s. deviation on
the mean. Again this is consistent with that derived from the
supergiant in this cluster and the kinematic distance.

In the case of Danks~2, the O6-8 stars all have $K$-band fluxes which
are consistent with them being dwarfs, when taking into account both
the kinematic distance and the spectrophotometric distance of
D2-1. These stars have an average spectrophotometric distance of
3.4$\pm$0.2kpc, consistent with G305's kinematic distance of
4.2$\pm$2.0\,kpc. The O8-9 stars however appear to be too bright to be
class {\sc v} stars, and have probably evolved away from the
main-sequence. There $K$-band brightnesses are more typical of giants,
though we consider the absolute brightnesses of class {\sc iii} stars
to be too uncertain to determine a reliable spectrophotometric
distance. 

Since the distances to the two clusters are within the errors of one
another, and the distance of Danks~2 is constrained by only two
measurements, for the rest of this paper we make the assumption that
the two are at the same distance. This is a reasonable assumption to
make, since if the size of the G305 cloud along the line-of-sight is
comparable to its angular size, the maximum difference in distances
between the two clusters is only $\sim$30pc. Taking the weighted
average of the spectrophotometric distances for the dwarfs only, as
well as the kinematic distance, we find a distance to the clusters of
3.8$\pm$0.6kpc.

\begin{table*}
  \centering
  \caption{Spectrophotometric distances of the stars in each cluster,
    assuming luminosity classes {\sc v}, {\sc iii} and {\sc i},
    denoted as D$_{\sc v}$, D$_{\sc iii}$, D$_{\sc i}$
    respectively. Calculations use the absolute $K$-band magnitudes of
    \citet{Martins-Plez06} and \citet{Crowther06}. }
  \begin{tabular}{lccccc}
    \hline \hline
Star & $m_{222}$ & Spec Type & D$_{\rm V}$/kpc & D$_{\rm III}$/kpc &
D$_{\rm I}$/kpc \\
\hline 
{\it Danks~1} \\
 D1-7 &  9.511 & O4-O6 &  {\bf3.21 $\leftrightarrow$  4.15}   &  4.59 $\leftrightarrow$  5.20   &  6.08 $\leftrightarrow$  6.11   \\ 
 D1-8 &  9.655 & O4-O6 &  {\bf3.43 $\leftrightarrow$  4.44}   &  4.91 $\leftrightarrow$  5.56   &  6.50 $\leftrightarrow$  6.53   \\ 
 D1-9 & 10.212 & O4-O6 &  {\bf4.43 $\leftrightarrow$  5.73}   &  6.34 $\leftrightarrow$  7.19   &  8.40 $\leftrightarrow$  8.44   \\ 
D1-10 & 10.581 & O6-O8 &  {\bf4.06 $\leftrightarrow$  5.25}   &  6.58 $\leftrightarrow$  7.52   &  9.96 $\leftrightarrow$  9.96   \\ 
D1-11 & 11.229 & O9-B3 &  {\bf1.19 $\leftrightarrow$  4.78}   &                  --             & 14.18 $\leftrightarrow$ 24.42   \\ 
D1-12 & 11.569 & O9-B3 &  {\bf1.39 $\leftrightarrow$  5.60}   &                  --             & 16.59 $\leftrightarrow$ 28.56 \smallskip \\ 
{\it Danks~2} \\
 D2-4 &  9.510 & O8-O9 &  2.37 $\leftrightarrow$  2.70   &  {\bf4.11 $\leftrightarrow$  4.39}   &  6.64 $\leftrightarrow$  7.02   \\ 
 D2-6 &  9.606 & O8-O9 &  2.47 $\leftrightarrow$  2.83   &  {\bf4.30 $\leftrightarrow$  4.58}   &  6.94 $\leftrightarrow$  7.33   \\ 
 D2-7 &  9.644 & O8-O9 &  2.52 $\leftrightarrow$  2.88   &  {\bf4.37 $\leftrightarrow$  4.67}   &  7.06 $\leftrightarrow$  7.46   \\ 
 D2-5 &  9.583 & O6-O8 &  {\bf2.80 $\leftrightarrow$  3.62}   &  4.53 $\leftrightarrow$  5.18   &  6.86 $\leftrightarrow$  6.87   \\ 
 D2-8 &  9.805 & O6-O8 &  {\bf3.10 $\leftrightarrow$  4.01}   &  5.02 $\leftrightarrow$  5.74   &  7.60 $\leftrightarrow$  7.60   \\ 
\hline 
\label{tab:specphot}
  \end{tabular}
\end{table*}

{\bf}

\subsection{Stellar populations and cluster ages} \label{sec:ages}

The stellar populations of the two clusters -- the presence of
early-to-mid O dwarfs and supergiants -- clearly indicate ages
$\la$6Myr. At these young ages it is not useful to simply fit model
isochrones to the near-IR colour-magnitude diagrams, as there is a
large degeneracy in age. Instead, we estimate ages for the two
clusters by three methods: analysis of their stellar populations;
examining the MS turn-offs; and also by studying the low-mass pre-MS
population of each cluster.

\subsubsection{Danks~1}
In the previous section we argued that the O4-6 stars in Danks~1 were
very likely on the main-sequence, whereas the stars which are slightly
brighter (the O6-8If, WNLh and OBI stars) appear to be post-MS
objects. From this information, we conclude that the MS turn-off
corresponds to a spectral-type of O4-6. From \fig{fig:phot1}, this
implies a ZAMS mass of 60$\pm$20\msun\ once the error in distance is
taken into account \citep{Mey-Mae00,Martins05}\footnote{We determine a
  scale of spectral-type versus mass using the mass-temperature scale
  defined by the stellar structure models of \citep{Mey-Mae00}, and
  the spectral-type -- temperature scale of \citet{Martins05}. }. The
MS lifetime for such stars is around 3-4Myr \citep{Mey-Mae00}, which
is therefore an upper limit to the age of Danks~1.

In the CMD of Fig.\ \ref{fig:phot1}, a departure from the ZAMS is
seen at magnitudes fainter than F222M$\sim$15. This can be interpreted
as the point at which lower mass stars in the cluster are beginning to
arrive on the MS. From the implied mass at which stars are on the MS,
we can get an independent measure of the cluster's age. The brightness
at which the stars are seen to join the MS corresponds to a mass of
4$\pm$1\msun. From comparison with models of pre-MS evolution
\citep[e.g.][]{P-S99,Siess00}, we find the age of the low-mass stellar
population of Danks~1 to be 1-2Myr.

The most massive stars present in the cluster can also be used as an
age discriminant, since the lifetime of a star is closely related to
its initial mass. The most massive stars in Danks~1 are likely to be
the three WNLh stars. Such stars are also present in the Arches
cluster, where their initial masses have been found to be as high as
120\msun\ \citep{Martins08}. Following Martins et al., we use their
mean $K$-band bolometric correction for WNLh stars, ${\it BC}_{K} =
-4.21\pm0.26$, as well as the average cluster extinction
(Sect.\ \ref{sec:extinct}) and distance modulus to G305
(Sect.\ \ref{sec:dist}). Using these numbers, we find that the
luminosities of D1-1, D1-2 and D1-5 are $\log (L/L_{\odot})$ = 6.5,
6.2, and 5.9 respectively (all $\pm$0.2dex). Clearly these three stars
are all intrinsically luminous, and therefore very massive. While
accurate spectrophotometric mass determinations await quantitative
modelling of their spectra, we can tentatively say here that the
masses of D1-1 and D1-2 are likely to be in excess of
90\msun\ \citep[comparing with Geneva rotating
  models,][]{Mey-Mae00}. This then places an upper limit to the age of
the cluster of 3Myr.

To summarize, an age of 1.5$^{+1.5}_{-0.5}$Myr for Danks~1 is
consistent with all pieces of evidence from both the high mass stars
and the low mass pre-MS stars. While it has been found that these two
age indicators can give contradictory results
\citep[e.g. Westerlund~1,][]{Brandner08}, and that pre-MS isochrones
may give ages which are systematically younger than those indicated by
MS stars \citep{Naylor09}, we find no evidence for such a discrepancy
here.

Finally, we mention a star in Danks~1 that does not seem to fit with
our derived age for the cluster. The star D1-3 has the appearance of a
`normal' blue supergiant, with narrow absorption lines, and the
absence of significant line emission implies a relatively weak
wind. It is therefore natural to conclude that this star is one of
moderate initial mass, say $\sim$20-40\msun, in an advanced
evolutionary state. This would imply an age of 4-10Myr, which is
clearly at odds with the other evidence from the cluster. This
suggests that D1-2 was not born with the rest of the stars in Danks~1,
and is instead part of a population of older stars which includes the
other evolved massive stars seen in the field of G305 (see Table
\ref{tab:otherstars}). It is possible that these stars formed along
with the older Danks~2 (see next section) and were dynamically
ejected. For a projected cluster separation of 40pc, if the star was
ejected from Danks~2 $\sim$3Myr ago then this implies a runaway
velocity of $\sim$15\kms, which is certainly not unreasonable.

%The
%implied runaway velocities of order $\sim$100\kms, though large, are
%not without precedent \citep[see e.g. the halo B~star
%  HD~271791,][]{Heber08}. Proper motion and/or quantitative modelling
%of D1-3 will be required to resolve the question of the star's nature
%and its relation to the G305 system.

%{\bf Ben - is this correct? I get much smaller velocities required to shift e.g. 
%the WCs from e.g. Dks 2 - e.g. 10km/s allows them to travel $\sim$7pc in roughly 0.7Myr 
%which seems emminently possible. Also, should this bit of the discussion fit in later (as I have it now?)}%

\subsubsection{Danks~2}
Using the same diagnostics of the cluster age, the evidence suggests
that Danks~2 is somewhat older than Danks~1. The earliest spectral
type MS stars are O6-8, consistent with masses of 30-40\msun\ and
hence an upper limit to the age of 4-6Myr. The point at which the
pre-MS stars join the MS is at a lower mass than in Danks~1 --
$\sim$2\msun, indicating an age of 3Myr with upper and lower limits of
2-10Myr due to the uncertainty in distance.

The most luminous stars in Danks~2 can again be used as an age
indicator, though the results are less conclusive than in Danks~1. In 
particular, carbon-rich WR stars should be present in  clusters with ages between
3-6Myr, according to rotating Geneva models \citep{Mey-Mae00}. The
yellow star, D2-2, lies to the left of the ZAMS track in the
colour-magnitude diagram of \fig{fig:phot2}, suggestive that the star
does not belong to the cluster.

Our inability to determine a precise spectral type for the OB
supergiant D2-1, the brightest star in Danks~2 in the near-IR, means
that it places only weak constraints on the cluster age. The $K$-band
bolometric correction of a O8-B3 supergiant is ${\it BC}_{K} = -3.93
\rightarrow -1.55$, according to \citet{Martins-Plez06} and
\citet{Crowther06}, which implies a bolometric luminosity of $\log
(L_{D2-1}/L_{\odot}) = 5.3 \rightarrow 6.3$ for the coolest and
hottest temperatures respectively. A luminosity for D2-1 closer to the
lower limit seems more likely, since more luminous stars do not tend
to have the spectroscopic appearance of `normal' blue supergiants, and
instead have strong emission lines owing to their dense winds.

From the above evidence, we suggest that the age of Danks~2 is
3$^{+3}_{-1}$Myr. The error is dominated by the uncertainty on the
cluster's heliocentric distance. However, we can be certain that the
two clusters are at the same distance, within a few
percent. Therefore, it seems very likely that Danks~2 is the older of
the two clusters, being created 1.5$^{+1.5}_{-0.5}$Myr before Danks~1.

\begin{figure*}
  \includegraphics[width=8.5cm,bb=5 10 710 560]{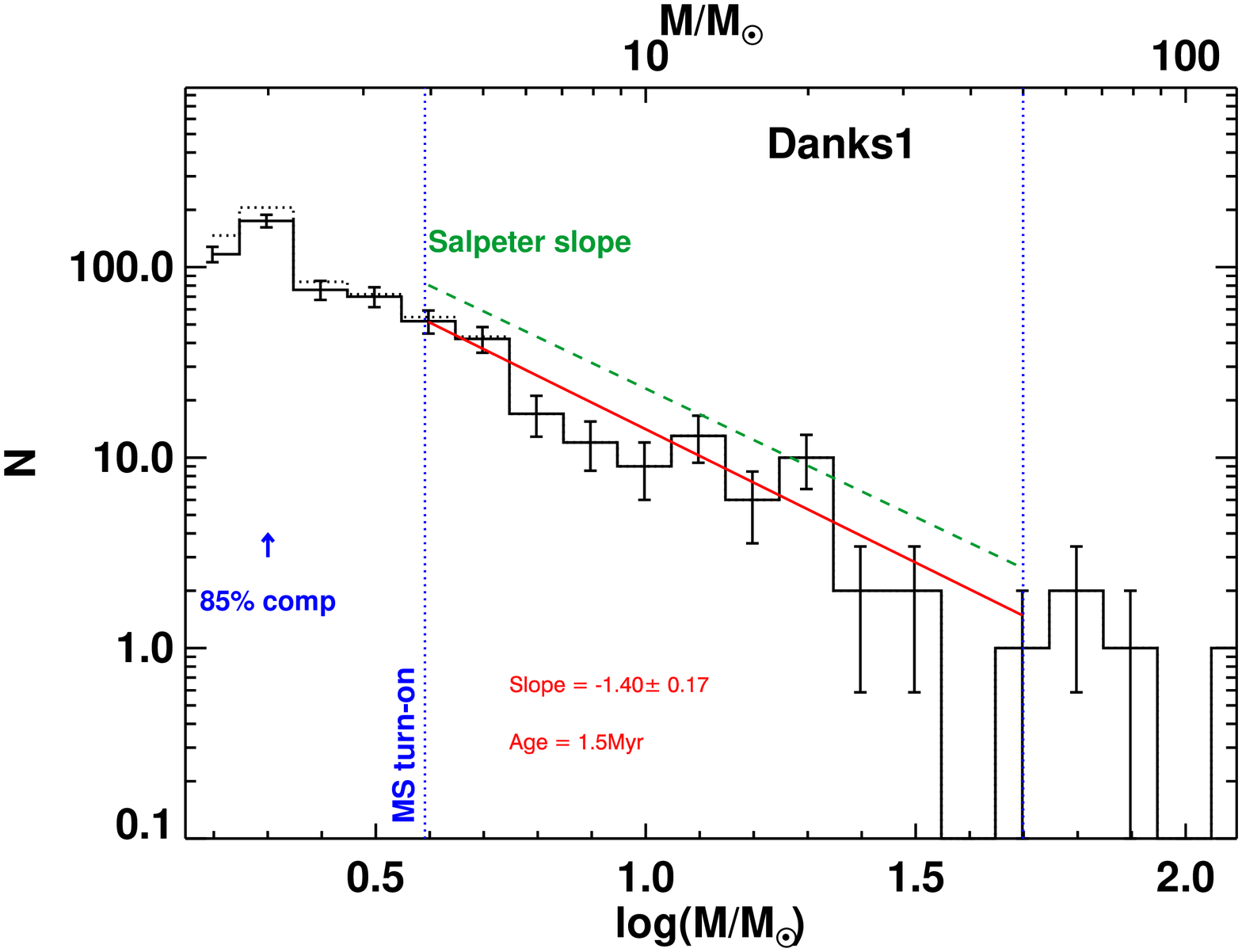}
  \includegraphics[width=8.5cm,bb=5 10 710 560]{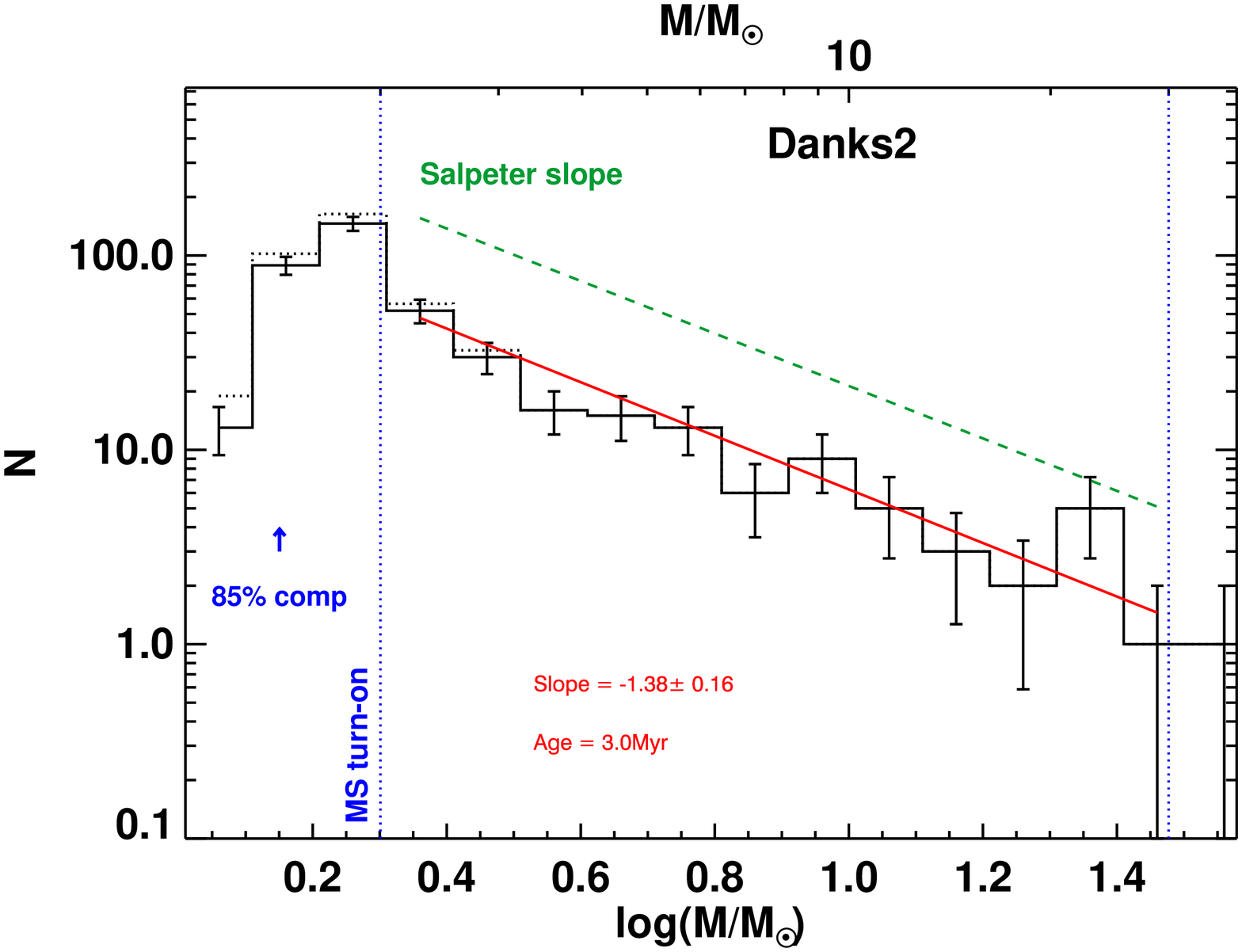}
  \caption{Initial Mass Functions (IMFs) for the two clusters: Danks~1
    ({\it left}) and Danks~2 ({\it right}). The solid and dotted
    histograms show the data before and after correction for
    completeness. The green dashed line illustrates the Salpeter slope. The
    slope of each mass function, shown as the red solid line, is
    computed between the MS turn-on and turn-off marked by the blue
    dotted lines. }
  \label{fig:mf}
\end{figure*}

\subsection{Initial mass functions}
In order to construct initial mass functions (IMFs) for the two clusters, we
must first create luminosity functions (LFs) which are decontaminated
of foreground stars. We experimented with two methods of doing
this. In the first method, we contructed LFs for both the cluster and
control fields, then subtracted one from the other once completeness
effects had been taken into account. Secondly, we used the F222M
magnitudes of the stars in the field-corrected colour-magnitude
diagrams of Figs.\ \ref{fig:phot1} and \ref{fig:phot2}. The
completeness corrections applied to these LFs were those of the
cluster fields, as these completeness limits are typically higher than
those of the control fields. In any event, the limit down to which we
chose to measure the IMFs is well above the point at which the
photometry becomes incomplete (see below). We also experimented with
de-reddening the stars in the decontaminated LF back onto the ZAMS
track, under the assumption that stars lying slightly to the right of
the track are cluster stars with extra local extinction. Any stars
which had (F160W-F222M) excesses greater than 0.6mags were discarded,
as these most likely belong to a separate background population,
though we note that without spectra we cannot
completely rule out that these objects are local to the clusters. 

To convert the LFs to IMFs, we use the evolutionary models of the
Geneva group. The latest versions of these take into account stellar
rotation \citep[e.g.][]{Mey-Mae00}, but computations do not exist for
stars with masses below 9\msun, since rotation does not have a large
impact on intermediate to low mass stars. For this reason we splined
together the rotating models with the older non-rotating models of
\citep{Schaller92} at 9\msun. As a check on the robustness of our
results, we also experimented with the older non-rotating models with
higher/lower metallicities, and varying mass-loss rates. Instrumental
magnitudes and colours were determined the same way as in
Sect.\ \ref{sec:ages}.

%using the calculations of
%\citep{Martins-Plez06} for the high-mass hot stars, and using the numbers
%compiled by \citep{Blum00} for the lower mass stars.

%we use the non-rotating evolutionary models
%of the Geneva group, which have instrumental colours available
%\citet{Schaller92}. For each cluster we used isochrones with the range
%of ages derived in \ref{sec:ages}, the distances of
%Sect.\ \ref{sec:dist}, and experimented with different metallicities. 

In Fig.\ \ref{fig:mf} we show the IMFs for the two clusters, derived
using the colour-magnitude corrected LFs, and using Solar-metallicity
isochrones including rotation at the ages indicated in each panel. In
each case, we measure the slope of the IMF {\it only in the mass
  ranges where we are sure that stars are on the main-sequence.} The
low mass stars which have not yet reached the MS will have colours and
magnitudes which are systematically different from those predicted by
main-sequence evolution codes; while the post-MS behaviour of
high-mass stars is extremely uncertain. The pre-MS mass limit is
measured from the kink in the CMD seen in both cluster sequences at
$m_{\rm F222} \approx 15$. The post-MS mass is determined from the
point at which the stellar spectrophotometric distances are no longer
consistent with luminosity class V stars. In practise, the upper limit
for the range in which we measure the IMF (dotted blue lines in
\fig{fig:mf}) is set by the point at which the number of stars per bin
drops below 2, where the uncertainties become non-poissonian.

In each cluster we see slopes that are consistent with the
Salpeter value of $\Gamma = -1.35$. Danks~1 has a slope $\Gamma =
-1.40 \pm 0.17$, while Danks~2 has a slope $\Gamma = -1.38 \pm
0.16$. These measurements are robust to the type of evolutionary model
used in the isochrones; the effect of varying mass-loss rate, stellar
rotation and metallicity is small compared to the uncertainty. Varying
the cluster age within the measured uncertainties has little impact,
since we measure the slope of the IMF only for objects on the MS where
stellar properties change very little over $\sim$10$^6$yrs. Finally,
varying the cluster distances between the upper and lower limits
(3.2-4.4kpc) produces changes in the slope which are small compared to
the uncertainties. We can therefore say that we find no significant
evidence for variations in IMF between the two clusters in G305, and
no significant deviation from the Salpeter slope.

If, as we have argued in Sect.\ \ref{sec:ages}, Danks~1 is in a
pre-supernova (SN) state, the IMF of this cluster may be used to investigate
the upper end of the IMF. Observations of the Arches cluster have
indicated that there may be an upper mass limit of
$\sim$150\msun\ \citep{Figer05}, though observations of R136 in the
Large Magellanic Cloud suggest that stars may form with masses greater
than this \citep{Crowther10}. Quantitative spectral modelling of the
brightest stars in Danks~1 would yield accurate bolometric
luminosities, and, by comparing to evolutionary models, initial
masses. It would then be possible to discuss these objects in the
context of the high-mass end of the IMF.

\subsection{Cluster masses}
In order to determine the initial masses of the two clusters, we
simply fit the observed IMFs measured in the previous section with a
functional form. Since all functional forms of the IMF have a slope
which is approximately Salpeter-like in the mass-range appropriate to
our measurements, we simply scale the functional IMF to fit our
observations and then integrate over all stellar masses. 

Fitting a Kroupa IMF to our observations \citep{Kroupa01}, we find
that the initial masses of Danks~1 and Danks~2 are
8000$\pm$1500\msun\ and 3000$\pm$800\msun\ respectively. The
uncertainties are due to a combination of errors carried forward from
cluster age and distance, as well as a stochastic error due to low
number statistics at the high-mass end.

We note that the Kroupa IMF puts a large amount of mass into stars
with sub-Solar masses, going all the way down to 0.01\msun. If we were
to use, for example, a Chabrier IMF \citep{Chabrier03}, we would find
that the inferred cluster masses would be approximately a factor of
two lower. 

\begin{figure}
  \centering
  \includegraphics[width=8cm]{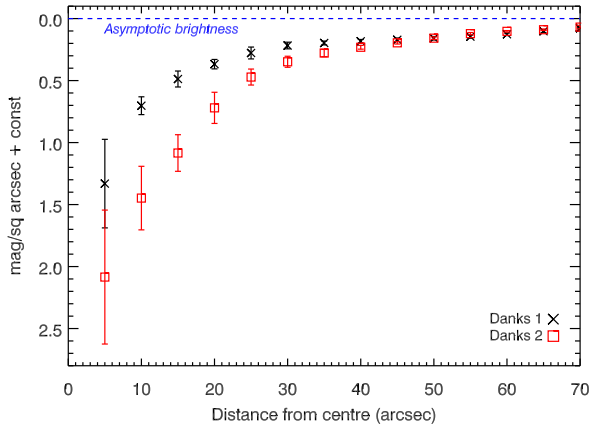}
  \caption{Radial surface brightness profiles for the two clusters.}
  \label{fig:radprofs}
\end{figure}

\begin{table*}
  \centering
  \caption{Summary of the clusters' physical properties}
  \begin{tabular}{lcccc}
    \hline
    Cluster & Age (Myr) & $M_{\rm init}$ (\msun) & R$_{0.5}$ (pc) &
    $\log (\rho /$\msun\,pc$^{-3}$) \\
    \hline
    \hline
    Danks~1 &    1.5$^{+1.5}_{-0.5}$    &  8000$\pm$1500  &  0.17$\pm$0.05 &
    5.5$^{+0.5}_{-0.4}$ \smallskip \\
    Danks~2 &    3.0$^{+3.0}_{-1.0}$    &  3000$\pm$800  &   0.36$\pm$0.09 &
    4.2$^{+0.5}_{-0.4}$ \\
    \hline
  \end{tabular}
  \label{tab:results}
\end{table*}

\subsection{Cluster sizes and densities}
We determined the half-light radius of each cluster $R_{0.5}$ by
measuring the cumulative surface brightness profiles (SBPs) as a
function of distance from the cluster centre. Since our NICMOS
observations have a field-of-view too small to determine at which
radius the SBPs fall to the ambient level, we used 2MASS images to
make these measurements. We defined the cluster centre as the centre
of the NICMOS fields-of-view, but since the cluster centre is not
easily defined we repeated the measurements with the centre offset in
RA and DEC by $\pm$3\arcsec. 

The cumulative SBPs of the two clusters are shown in
\fig{fig:radprofs}. The error bars on each point represent the effect
of varying the position defined as the cluster centre. The half-light
radii of the clusters are given in Table \ref{tab:results}, where the
errors on these values take into account both the error on $R_{0.5}$
and the uncertainty in distance. Also summarized in Table
\ref{tab:results} are the cluster densities, ages and intergrated
masses.

%%%%%%%%%%%%%%%%%%%%%%%%%%%%%%%%%%%%%%%%%%%%%%%%%%%%%%%%%%%%%%%%%%%%%
%%%%%%%%%%%%%%%%%%%%%%%%%%%%%%%%%%%%%%%%%%%%%%%%%%%%%%%%%%%%%%%%%%%%%

\section{Discussion}

Our photometric and spectroscopic analysis of Danks 1 \& 2 reveal that
they are both rather massive, young and compact, with a total mass of
$\ga 10^4$\msun, and both forming over a short interval of
$\sim$2~Myr. In the following section we discuss how these clusters --
and by extention the G305 complex -- compare to other Galactic and
extra-galactic star-forming regions.

\subsection{Cluster comparison}

With a mass of $\sim$8000~M$_{\odot}$, an age of $\sim$1.5~Myr and a
stellar population containing both WNLh and early-mid O-type
supergiants, Danks~1 closely resembles a number of other massive young
clusters which are also associated with Giant H\,{\sc ii} regions.
These include Trumpler~14 \citep[located within the Carina
  nebula;][]{Smith06} and NGC~3603 \citep{Harayama08}, as well as the
Arches cluster \citep{Figer02}, which, given its location within the
central 50~pc of the Galaxy, cannot be uniquely associated with a
natal birthcloud.

By comparison, Danks~2 is both older, less massive and less dense than
the above clusters. Its mass and stellar population (and hence age)
are reminiscent of NGC~6231 at the centre of Sco~OB1 \citep{Sung98},
or Cl~1806-20 \citep{Figer05,Bibby08}, though the latter cluster does
not have any obvious associated GMC.

 The difference in densities between Danks~1 and
Danks~2 may be related to the dynamical evolution of young massive
star clusters. The Central Cluster aside, Danks~1, the Arches, and
NGC~3603 all have ages which are thought to be below 3Myr, and so all
may be yet to experience a SN of a cluster member. The
removal of mass from the cluster by SNe may leave the cluster
super-virial, causing it to expand \citep{Hills80,G-B06}. If Danks~2
has already experienced SNe, this may explain the difference in
central density between the two clusters.

\subsection{Cluster complex comparison}

If Danks~1 closely resembles both Trumpler~14 and NGC3603, one might
also ask whether the properties of the associated star forming regions
are also comparable?  At this point it is instructive to extend this
comparison to 30 Doradus and its central cluster R136 in the Large
Magellanic Cloud, the most luminous Giant H\,{\sc ii} region in the
Local Group.

\subsubsection{Multiple stellar populations}

The evidence presented here suggests that star formation activity
within the G305 commenced within the last 6~Myr. Following identical
arguments to those advanced for the W51 complex by \citet{Clark09},
the apparent lack of luminous Red Supergiants (RSGs) within the
complex - which would clearly be visible at IR wavelengths - is
consistent with the picture that star formation was absent beyond
$\sim$6~Myr ago.

At least two further populations of massive (post-)MS stars are
present within G305, with significant evidence for more recent and
ongoing massive star formation. \citet{Leistra05} find a mid-O
supergiant and two late-O/early-B main sequence objects located within
the cluster they designate as G305+00.2. Found within a small bubble
on the periphery of the complex, Hindson et al. (in prep.)  suggest an
age of $<$1~Myr for the compact H\,{\sc ii} regions associated with
it. 

Finally, a substantial diffuse population of massive stars appears to
be present, with \citet{Shara09} and \citet{Mauerhan11} discovering
additional WRs within the confines of the complex in addition to the
members of Danks~1 \& 2, WR~48a and the candidate dusty WCL
MSX305.4013+00.0170 (Table \ref{tab:otherstars}). Indeed this WR
population is significantly larger than that found within both Danks~1
\& 2 combined, while the presence of four WC stars implies a minimum
age of at least 3~Myr for this diffuse stellar component.

%Three
%possible scenarios may account for this population: (i) the {\em bona
%  fide} formation of isolated massive stars {\em in situ}, (ii) their
%formation in Danks~1 and/or 2 and subsequent ejection due SNe kicks
%and/or dynamical interactions and (iii) their formation in one or more
%clusters which have subsequently dissolved (via so called `infant
%mortality').

%Unfortunately, our current inability to distinguish
%between these options precludes us determining whether they are the
%`tip of the iceberg' of a third, hirtherto undetected massive stellar
%population; given simple number counts {\em if} they formed in
%clusters with a standard IMF they would likely herald the presence of
%a further $\sim$10$^4$M$_{\odot}$ of stars within the complex.

%The spectral types
%of these field stars indicate ages of $\ga$3~Myr.  The origin of this
%population is uncertain, as we are currently unable to distinguish
%between {\em in situ} formation, possible in conjunction with an
%associated cluster which has since dissolved, or their formation and
%subsequent ejection from either cluster as a result of dynamical
%interaction or a supernova in a binary \citep{Eldridge11}.

%\footnote{2MASS J13133230-6240125 (=
%  [SMG2009] 845-34; WC8), 2MASS J13122766-6244220 (= [SMG2009] 845-35;
%  WC7) and 2MASS J13124535-6305520 (= [SMG2009] 847-8;
%  WN6). \citet{Mauerhan09} likewise report a number of additional
%  candidate emission line objects with 5' of Danks 1 \& 2, although
%  with an absence of co-ordinates it is not possible to cross
%  correlate these with known stars.  } 

The NGC3603 and 30 Dor and Carina star forming regions also all appear
to host multiple stellar populations. Dominated by the young massive
cluster NGC~3603 (1-2~Myr), the presence of the blue supergiants Sher
23 \& 25 reveals an older ($\sim$4~Myr) population
\citep{Melena08}. \citep{Walborn-Blades97} likewise report a number of
different stellar groups within 30 Dor; (i) the central cluster R136
(2-3~Myr), (ii) an older ($\sim$4-6~Myr) population distributed across
the region, (iii) Hodge 301; a $\sim$10~Myr cluster to the NW of R136
(iv) the R143 association in the south-east ($\sim$6-7~Myr) and (v) a
very young ($\sim$1~Myr) population on the periphery of the complex,
which we return to below. Lastly, \citet{Smith06} summarise the
properties of Carina, again reporting a range of ages for the clusters
located within it (Tr14 at $\sim$1-1.5~Myr, Tr15 at 6$\pm$3~Myr, Tr16
at 2-3~Myr, Bo10 at $\sim$7~Myr and Bo11 at $<$3~Myr). \citet{Smith06}
also comments on the fact that in contrast to the previous examples,
these populations are rather widely distributed over the complex.
While the richest clusters (Tr14 \& 16) are centrally located the
remainder are observed at significant distances ($>$10~pc)in the outer
regions of the complex.  An age spread amongst the stellar populations
within the W51 complex is also apparent \citep{Clark09}, although in
this case there is no evidence for a compact central cluster, with
star formation apparently distributed throughout the host GMC.

\subsubsection{Ongoing star formation in cluster complexes}

As reported by \citet{C-P04}, \citet{Clark11} and \citet{Hindson10},
there is compelling evidence for a further generation of massive stars
forming with the G305 complex. Methanol masers and ultracompact
(UC-)H\,{\sc ii} regions, both unambiguous indicators of ongoing
massive star formation, are present within G305, being predominantly
located on the periphery of the bubble. Integrated IR- and
radio-fluxes likewise argue for the presence of a significant
population of embedded massive YSOs, while sub-mm observations reveal
the presence of a substantial reservoir
($\sim$6$\times$10$^5$~M$_{\odot}$) of cold molecular gas available to
fuel future activity.

The star forming complex associated with NGC~3603 also shares these
properties, with \citet{Nurnberger-Stanke03} describing the presence
of methanol massers, dusty embedded sources and molecular cores within
the confines of the remnant natal GMC.  Both IR- and radio-fluxes are
likewise comparable to those of G305 to within a factor of a $\sim$few
\citep[e.g.][]{Crowther-Conti03}.  A comparable morphology is also
observed for 30~Dor, with active star formation located on the
periphery of the cavity surrounding R136
\citep{Walborn-Blades97,Walborn02}. Finally, while such spatial
segregation between pre- and (post-)Main Sequence stars is less
apparent within Carina, the presence of ucH\,{\sc ii} regions
\citep{Brooks01} points to onoing massive stars formation,
particularly in the `Southern Pillars' region of the complex
\citep{Rathborne04}. Moreover, both integrated IR- \& radio-fluxes are
directly comparable to those of G305, as is the mass of cold molecular
material \citep{Smith-Brooks07}.

In all four cases it has been suggested that the complex morphologies
reflect the propogation of triggered star formation through the
molecular cloud. This impression is particularly strong for G305,
where the active star forming regions appear restricted to the
periphery of the cavity, and indeed with a `third generation' of
masers and ucH\,{\sc ii} regions associated with the bubble
surrounding the cluster G305+00.2 \citep{C-P04,Hindson10}. While such
morphologies are necessary, they are not considered sufficient
evidence for triggered star formation since it might be supposed that
the radiation- or wind blown-bubble is simply uncovering exisiting
activity. In this regard our unbiased multiwavelength datasets,
encompassing the whole G305 complex \citep{Clark11} will be
particularly valuable in distinguishing between these possibilities
via the identification of (massive) YSOs.  If star formation was
sequential, one would expect to find YSOs interior to the cavity but
none outside, whereas they should be distributed throughout the GMC if
this was not the case \citep[cf.][]{Smith10}. Indeed, our dataset will
allow us to constrain not only the propagation of star formation
through the GMC, but also the relative stellar yields of each
successive generation, a particularly interesting prospect given that
\citet{Smith10} suggest that this decreases with time (and successive
generations) in Carina.

%\footnote{
%  Given the differences in ages, an intriguing possibilityis that
%  Danks~2 triggered the formation of Danks~1.  To explore this
%  possibility further, we calculated the predicted number of SNe as a
%  function of time in the G305 region. This was done using isochrones
%  constructed from the \citet{Mey-Mae00} evolutionary tracks, and two
%  arrays of stellar masses sampled from a Kroupa IMF with total masses
%  equal to those of the two clusters. For isochrones of increasing age
%  we determine the maximum remaining stellar mass, and assume that any
%  cluster star with a greater mass has exploded as a SN. The
%  experiment was repeated 100 times in order to compensate for
%  stochastic variations. We found that in $\sim$40\% of the trials a
%  cluster of Danks~2's mass would contain a star massive enough to
%  have gone SN within 2.5Myr of its formation (i.e. mass
%  $\ga$100\msun). A SN blastwave travelling at $\sim$1,000\kms\ would
%  reach the current location of Danks~1 in only a few thousand
%  years. In addition to any SN ejecta, the most massive stars of the
%  cluster would have provided a significant source of ionising
%  radiation and mechanical energy (these properties will be quantified
%  in a future paper). Therefore, an age difference between the
%  clusters of 2.5Myr would be consistent with one triggering the
%  other. In Sect.\ \ref{sec:ages} we found that the age difference
%  between the two clusters is 1.5$^{+1.5}_{-0.5}$Myr, and so it is
%  entirely possible that the feedback from Danks~2 triggered the
%  star-forming event that created Danks~1.}.

Despite having a comparable cloud mass (Parsons et al., accepted with
minor revision) and massive star formation having been underway for
several million years \citep[cf.\ radio- and IR-fluxes similar to the
  above regions;][]{Conti-Crowther04}, it is interesting that the W51
complex appears not to host such triggered star formation.  Instead,
multiple, causally unconnected, star forming regions are found
throughout the cloud, albeit possibly synchronised by the presence of
an external agent such as the passage of a Galactic spiral density
wave \citep{Clark09,Kumar04}.  Nevertheless, the physical endpoint in
terms of the production of multiple young massive clusters with an age
range of $\la 10^6$ years appears remarkably similar to the
above complexes.

\subsection{A template for star formation}

Irrespective of the mechanism of star formation (multiseeded versus
triggered), these complexes present a template of how a
10$^6$-10$^7$M$_{\odot}$ GMC is converted into stars.  The process is
distributed in both time and space, yielding a number of rather
massive (10$^3$-10$^4$M$_{\odot}$) clusters with a significant age
spread over a region spanning an angular diameter of $\sim$30pc. Of
immediate interest is the fact that the properties of both NGC~3603
and Danks~1, in terms of integrated mass, density, age and stellar
content is directly comparable to that of the Arches.  Clearly the
production of such clusters is not dependent on the extreme conditions
of the Galactic centre, although with 3 clusters of similar masses (a
few $\times$10$^4$\msun; the Arches, Quintuplet and Galactic Centre)
the star formation rate here appears to have been larger over the past
few Myr than in the NGC~3603 and G305 complexes.

%Moreover, the presence of an addition, {\em non-clustered} population
%of massive stars within such complexes also appears to be ubiquitous,
%being seen for NGC3603, Carina, 30 Dor and even suggested for the
%Galactic Center \citep{Mauerhan10}. Although uncertain, the origin of
%these stars is of particular interest since they may either herald a
%non clustered mode of massive star formation or indicate the rapid
%dissolution of addition young clusters and hence a large star
%foramtion rate than inferred solely from the clusters.

Globally, these giant star-forming regions are also of interest due to
the fact that they mirror the structure of extragalactic star cluster
complexes, albeit with cluster masses several orders of magnitude
smaller. Indeed the integrated spectral energy distribution of Carina
is remarkably similar to those of Ultraluminous IR Galaxies
\citep{Smith07,Sanders-Mirabel96}, suggesting that a similar mode of
star formation is also present in such galaxies. Because of this,
determining the energy budget of Galactic examples such as Carina and
G305 is important since it will permit a calibration of the star
formation rate from the emergent flux.

While an accurate determination of the energy budget of G305 requires
both accurate stellar parameters and the mid-far IR flux (Clark et
al. in prep.), we simply note here that the current feedback from
Danks~1 \& 2 is likely to be dominated by the 3 WNLh stars. Utilising
the Lyman flux estimates for similar WNLh stars in the Arches cluster
scaled to the luminosities of these objects \citep{Martins08}, the
amount of ionizing radiation emitted by the three WNLh stars in
Danks~1 alone appears comparable to the total ionizing flux within the
entire G305 region estimated by \citet{C-P04}. This suggests that, as
with Carina, G305 suffers significant photon leakage. Hence,
estimating the stellar contents and star formation rates of such
regions using either IR or radio fluxes alone may result in
significant underestimates for all but the very youngest complexes,
where feedback has yet to uncover embedded clusters.

\subsection{Future evolution}

Finally we turn to the long term evolution of the G305 complex. As
stellar evolution drives mass loss, the long term survivability of the
clusters (Danks~1 \& 2 and G305+00.2) is somewhat uncertain
\citep{G-B06}. If they do disperse rapidly, G305 will increasingly
resemble a classical OB association, noting that the stars within such
aggregates appear to have formed over a comparable period
\citep[e.g. Cyg OB2 \& Sco OB1; Clark et
  al. submitted;][]{Negueruela08}. However, at present Danks~1 \& 2
appear to be tightly bound, with ratios of ages to crossing times of
order $\sim$10, suggesting that they will survive as clusters for
several million years \citep[see][]{Gieles-PZ11}. In this case, an
obvious point of comparison is the Perseus complex, which contains two
distinct star clusters, $h$ and $\chi$ Persei, surrounded by a halo of
stars of similar age. The complex has been recently studied in detail
by \citet{Currie10}, finding remarkably similar ages and masses for
the two clusters, 3700-4200\msun\footnote{These authors assumed a
  Miller-Scalo IMF to determine the total masses of the populations in
  the Perseus complex.}  and 14$\pm$1Myr, and an age for the halo
population which was only marginally younger, 13.5$\pm$1Myr.

A quantitative comparison with our results presented here show that
this is somewhat different to what we see in G305. The age spread we
find for Danks~1 and Danks~2 is small, 1-2~Myr, and 
will become more difficult to detect as the clusters approach the
same age as $h$ and $\chi$ Persei. However, we would still expect to
be able to detect a significant age difference between the central
clusters and the surrounding halo of stars. It is clear that the
remains of the G305 cloud is still forming stars today, implying an
age difference of up to 5~Myr. Therefore we may still expect to detect
a difference in the ages of the clusters and the surrounding halo when
the clusters are well beyond 20~Myr old.

An alternative comparison might be with the association of RSG
dominated clusters at the base of the Scutum Crux arm, which appear to
have formed over an extended period between $\sim$12--20~Myr ago
\citep{RSGC1paper,RSGC3paper}. In particular, RSGC3 would appear to
form a particularly good comparator, being surrounded by at least one
further lower mass cluster and a number of isolated RSGs of similar
ages and masses within a $\sim$30~pc radius
\citep{RSGC3paper,RSGC5paper}.  While currently separate, with a
(projected) separation of only $\sim$4pc a combination of cluster
expansion driven by stellar mass loss and dynamical interaction could
lead to the effective merger of Danks~1 \& 2, further emphasising the
similarity between the 2 regions.

%; indeed the formation of massive
%clusters by merger has previously been suggested by Fellhauer \&
%Kroupa (\cite{fellhauer}) and Kroupa \cite{kroupa}).

In the case of Danks 1\& 2 if this were to occur before the RSG
dominated phase ($>$10~Myr) this would lead to a mixture of stellar
spectral types not predicted by current evolutionary theory. A similar
explanation has been proposed to explain the stellar population of the
young ($\sim$7Myr) massive cluster NGC~1569-A.The integrated spectrum
displays features attributed both to Red Supergiants (RSGs) and WRs
\citep{G-D97,Hunter00}, which stellar population models struggle to
produce simultaneously for a single instantaneous starburst. Analysis
of the point-spread function and velocity dispersion has also
indicated two distinct components to NGC~1569-A, now labelled A1 and
A2 \citep{DeMarchi97,G-G02}. Further, \citet{Maoz01} found that the WR
spectra signatures were confined to A2. The current understanding of
NGC~1569-A1 and NGC~1569-A2 is that they have comparable masses of
$\sim 6 \times 10^5$\msun, but ages of $\la$5Myr and $\sim$8Myr
respectively; a comparable age spread to that found between Danks~1
and Danks~2, though the NGC~1569-A clusters are slightly older and
orders of magnitude greater in mass.

\section{Summary and future work}
We have provided the first comprehensive study of the two central
clusters of the G305 star-forming complex, Danks~1 and Danks~2. We
determine a distance to the clusters and the host molecular cloud of
3.8$\pm$0.6kpc. This takes into account both the spectrophotometric
and kinematic measurments, which are in agreement with each other. The
total stellar masses of Danks~1 and Danks~2 are
8000$\pm$1500\msun\ and 3000$\pm$800\msun\ respectively. Analysis of
their initial mass functions (IMFs) shows that both are consistent
with Salpeter. A further population of apparently isolated massive
stars, predominantly comprising WRs, is also present within the
confines of the complex.

From analysis of the stellar content and pre-main-sequence stars of
the two clusters, we have been able to piece together the star-forming
history of the G305 complex. Danks~2 is the oldest object in the
complex, with an age of 3$^{+3}_{-1}$Myr. The younger age of Danks~1,
1.5$^{+1.5}_{-0.5}$Myr, is consistent with its formation being
triggered by the feedback of Danks~2. It is likely that the combined
winds from the two clusters are responsible for the evacuation of the
complex's central cavity, and perhaps are responsible for the
considerable amount of star formation that has occurred in the last
$\sim$0.5Myr around the periphery of the G305 cloud. The origin of the
diffuse population of evolved massive stars is less clear, though the
presence of both WN and WC stars is consistent with this population
forming at a similar time to the two central clusters. 

%an origin in
%either Danks~1 \& 2. Alternatively, if not runaways, they could either
%have formed {\em in situ}, via an isolated instead of clustered mode
%of star formation, or in a cluster(s) subsequently lost to infant
%mortality; if correct, the lattter scenario would imply a
%significantly higher rate of star formation for the G305 complex.

%PUT IT ALL INTO CONTEXT....

The masses and stellar content of the two clusters, combined with the
significant numbers of Wolf-Rayet stars in close vicinity to the
clusters, and the numerous massive protostars in the surrounding
molecular cloud, make G305 one of the most bountiful regions for
massive star formation known in the Galaxy, comparable to the Carina
nebula. As such, the region represents the perfect template for
calibrating unresolved observations of similar regions at
extra-galactic distances.

In terms of work for the future, the high-quality spectra presented
here will be modelled with stellar atmosphere codes. This will allow a
quatitative estimate of the amount of feedback from the clusters, both
in terms of the ionizing radiation and the mechanical energy from
stellar winds. Also, it will provide an estimate of the bolometric
luminosities for the most massive stars in the cluster, and the
universality of the IMF up to the highest stellar masses.

\section*{Acknowledgments}

We thank Fabrice Martins for supplying us with the spectra of stars in
the Arches cluster. We also thank the anonymous referee whose
comments and suggestions helped improve the paper. BD is supported by
a fellowship from the Royal Astronomical Society. This work is in part
based on observations made with the NASA/ESA Hubble Space Telescope,
obtained at the Space Telescope Science Institute, which is operated
by the Association of Universities for Research in Astronomy, Inc.,
under NASA contract NAS 5-26555. These observations are associated
with program \#11545. Support for program \#11545 was provided by NASA
through a grant from the Space Telescope Science Institute, which is
operated by the Association of Universities for Research in Astronomy,
Inc., under NASA contract NAS 5-26555. This work is in part based on
observations collected at the European Organisation for Astronomical
Research in the Southern Hemisphere, Chile, under programme number
077.C-0207(B). Financial support from the Spanish Ministerio de Ciencia e Innovaci\'on under projects AYA2008-06166-C03-02 and AYA2010-21697-C05-01 is acknowledged. 

\bibliographystyle{/fat/Data/bibtex/apj}
\bibliography{/fat/Data/bibtex/biblio}

\end{document}